\documentclass[aps,prb,superscriptaddress,showpacs,floatfix,twocolumn]{revtex4-2}
\usepackage[utf8]{inputenc}

\usepackage[tbtags]{mathtools}
\usepackage{graphicx}
\usepackage{xcolor}
\usepackage{siunitx}
\usepackage{dsfont}
\usepackage{multirow}
\usepackage{braket}
\usepackage{hyperref}
\usepackage[capitalize]{cleveref}

\usepackage[a4paper, margin=2cm]{geometry}

\newcommand{\etalNoPeriod}{\textit{et~al}}
\newcommand{\etal}{\etalNoPeriod{}.\@}

\DeclarePairedDelimiter{\abs}{\lvert}{\rvert}
\DeclarePairedDelimiter{\paren}{\lparen}{\rparen}

\DeclareMathOperator*{\argmin}{argmin}
\renewcommand{\vec}[1]{\mathbf{#1}}

\DeclareSIUnit\angstrom{\text {Å}}

\newcommand{\eul}{\mathrm{e}}
\newcommand{\ci}{\mathrm{i}}
\newcommand{\id}{\mathds{1}}
\newcommand{\nk}{{n \vec{k}}}
\newcommand{\nkp}{{n' \vec{k}'}}
\newcommand{\mass}{M}
\newcommand{\ev}{\varepsilon}
\newcommand{\plka}{{\partial_{l \kappa \alpha}}}

\newcommand{\bzint}{{\int_\text{BZ}}}
\newcommand{\BZvol}{{\Omega_\text{BZ}}}
\newcommand{\kint}[1]{{\frac{\mathrm{d}^3{#1}}{\BZvol}}}
\newcommand{\paw}[1]{\tilde{#1}}
\newcommand{\pawT}{\hat{\mathcal{T}}}
\newcommand{\ppsi}{\paw{\psi}}

\newcommand{\grid}[3]{\ensuremath{#1{\times}#2{\times}#3}}
\newcommand{\gridc}[1]{\grid{#1}{#1}{#1}}

\begin{document}

\title{Zero-point Renormalization of the Band Gap of Semiconductors and Insulators Using the PAW Method}

\author{Manuel Engel}
\email[]{manuel.engel@univie.ac.at}
\affiliation{University of Vienna, Faculty of Physics and Center for Computational Materials Physics, Kolingasse 14-16, A-1090 Vienna, Austria}

\author{Henrique Miranda}
\affiliation{VASP Software GmbH, Sensengasse 8/12, A-1090, Vienna, Austria}

\author{Laurent Chaput}
\affiliation{LEMTA - Université de Lorraine, CNRS, UMR 7563, 2 avenue de la Forêt de Haye, 54518 Vandœuvre Cedex, France}

\author{Atsushi Togo}
\affiliation{Research and Services Division of Materials Data and Integrated System, National Institute for Materials Science, Tsukuba, Ibaraki 305-0047, Japan}

\author{Carla Verdi}
\author{Martijn Marsman}
\author{Georg Kresse}
\affiliation{University of Vienna, Faculty of Physics and Center for Computational Materials Physics, Kolingasse 14-16, A-1090 Vienna, Austria}

\begin{abstract}
    We evaluate the zero-point renormalization (ZPR) due to electron-phonon interactions of 28 solids using the projector-augmented-wave (PAW) method.
    The calculations cover diamond, many zincblende semiconductors, rock-salt and wurtzite oxides, as well as silicate and titania.
    Particular care is taken to include long-range electrostatic interactions via a generalized Fröhlich model, as discussed in Ref.~\cite{verdi-polar,sjakste-polar}.
    The data are compared to recent calculations~\cite{miglio-zpr} and generally very good agreement is found.
    We discuss in detail the evaluation of the electron-phonon matrix elements within the PAW method.
    We show that two distinct versions can be obtained depending on when the atomic derivatives are taken.
    If the PAW transformation is applied before taking derivatives with respect to the ionic positions, equations similar to the ones conventionally used in pseudopotential codes are obtained.
    If the PAW transformation is used after taking the derivatives, the full-potential spirit is largely maintained.
    We show that both variants yield very similar ZPRs for selected materials when the rigid-ion approximation is employed.
    In practice, we find however that the pseudo version converges more rapidly with respect to the number of included unoccupied states.
\end{abstract}

\maketitle

\section{Introduction}

Electron-phonon interactions are among the most relevant processes governing the temperature dependence of the electronic properties of materials.
Accurately predicting and controlling these properties is crucially important for the development of new technologies~\cite{batteries-1,batteries-2,solar-cells-1,solar-cells-2,organic-electronics,thermoelectrics-1,thermoelectrics-2,thermoelectrics-3,thermoelectrics-4}.
For instance, electron-phonon interactions are essential for explaining the temperature-dependent magnitude of the band gap of insulators and semiconductors, as demonstrated repeatedly~\cite{marini-ab-initio-2008,ponce-temperature-2014,zacharias-one-shot-2016,zacharias-one-shot-2020}.
Even at absolute zero temperature, these processes can modify the band gap significantly, which is commonly referred to as a zero-point renormalization.

Historically, the first ZPR calculations for real materials were performed by Allen, Heine and Cardona (AHC)~\cite{allen-heine,allen-cardona} around 1980.
Their method, based on second-order Rayleigh-Schrödinger perturbation theory~\cite{rayleigh-perturb,schroedinger-perturb}, forms the basis for many modern approaches.
Over the years, there have been a number of theoretical and computational advances that generalize or improve upon AHC theory.

Firstly, AHC theory only considers static, non-frequency dependent perturbations.
As a result, the expression for the ZPR in AHC theory does not account for the energy transfer from the electronic to the ionic subsystem during phonon emission and absorption.
This approximation is often referred to as the adiabatic AHC theory.
A non-adiabatic formulation can be derived from time-dependent or many-body perturbation theory~\cite{giustino-review}.
Inclusion of the energy transfer also avoids numerical divergence problems in the case of polar materials in the limit of small phonon wave vectors~\cite{ponce-static-vs-dynamic,zacharias-one-shot-2020}.

In 2007, Giustino~\etal{}~\cite{giustino-wannier-interpol-1,giustino-wannier-interpol-2} introduced an interpolation scheme for electron-phonon matrix elements based on maximally localized Wannier functions~\cite{marzari-mlwf-1,marzari-mlwf-2}.
This removed the need for often prohibitively expensive density-functional perturbation-theory (DFPT) calculations on dense q-point grids.
Over time, this and related interpolation techniques have successfully been applied to the electron-phonon problem~\cite{giustino-diamond,calandra-interp,li-elph-linear-interp,eiguren-wannier-2008,brunin-elph-abinit,chaput-elphon,Agapito-elph-atomic-orbitals,engel-elph-paw}.

In order to properly treat polar materials using interpolation techniques, the long-range behavior of the electron-phonon interaction ought to be accounted for explicitly.
To this end, early considerations by Fröhlich~\cite{froehlich-elph} and Vogl~\cite{vogl-elph} were generalized to develop a long-range electron-phonon potential from first principles that accurately captures the dipole interaction~\cite{verdi-polar,sjakste-polar}.
Contributions from the quadrupole interaction were recently covered by Brunin~\etal{}~\cite{brunin-quadrupole} and Jhalani~\etal{}~\cite{jhalani-quadrupole}.

An extension of adiabatic AHC theory to the PAW method~\cite{bloechl-paw,kresse-paw} was provided by Engel~\etal{}~\cite{engel-elph-paw}.
Since the PAW method accounts for the exact shape of the all-electron (AE) wave functions, it should improve the accuracy compared to traditional pseudopotential methods, yet largely retains the computational performance of the latter.
An alternative derivation was provided by Chaput~\etal{}~\cite{chaput-elphon}, relying on a rigorous full-potential formulation of the electron-phonon matrix element expressed within the PAW framework.
This even applies in the non-adiabatic case, i.e., when energy is transferred during phonon emission and absorption.

Alternative supercell-based approaches to calculate the band-gap renormalization include Monte-Carlo simulations~\cite{monserrat-mc,patrick-mc}, molecular dynamics~\cite{kundu-md-2021,zacharias-md} and other adiabatic supercell-based approaches.
The latter rely on one or more specially chosen, frozen atomic displacements~\cite{monserrat-supercell-1,antonius-supercell,monserrat-supercell-2,zacharias-one-shot-2016,karsai-one-shot,zacharias-one-shot-2020,monserrat-thermal-lines-2016}.
Although these methods have advantages such as inclusion of selected higher-order terms, they also have the disadvantage that they usually remain in the adiabatic regime (neglect of energy transfer during emission and absorption).
Furthermore, for polar materials it can be very difficult to reach the required supercell convergence.
On the other hand, they are easily applicable to methods beyond density-functional theory (DFT), so that studies comparing DFT and the GW method became possible recently~\cite{karsai-one-shot}.
A comprehensive review of electron-phonon physics from the point of view of first-principles calculations was given by Giustino~\cite{giustino-review}.

In this work, we present results for the band-gap ZPR of various semiconductors and insulators within the non-adiabatic AHC theory.
The calculations are performed within the framework of DFT using the PAW method and are based on the computational approach proposed by Chaput~\etal{}~\cite{chaput-elphon} and Engel~\etal{}~\cite{engel-elph-paw}. 
The present implementation relies partly on the VASP code~\cite{vasp-1,vasp-2,vasp-3,vasp-4} but also uses an external program to calculate the derivatives of the self-consistent Kohn-Sham (KS) potential using supercells.
Importantly, two different equations for the electron-phonon matrix elements are compared and assessed, namely a pseudized~\cite{engel-elph-paw} and a full-potential~\cite{chaput-elphon} formulation of the PAW matrix element.
Although they describe distinctively different electron-phonon matrix elements, both formulations are shown to yield identical ZPR under certain conditions.
The details can be found in \cref{app:ae-ps-equiv-proof}.

The general theory and methodology underpinning this work is presented in \cref{sec:methods}.
In \cref{sec:ahc-theory}, a brief summary of non-adiabatic AHC theory is given and the ZPR is expressed in this framework.
In \cref{sec:elphon-paw}, the calculation of the ZPR is discussed in the context of the PAW method.
A short description of the computational workflow is provided in \cref{sec:comp-approach}.
The results are presented and discussed in \cref{sec:results} followed by our conclusions in \cref{sec:conclusion}.

\section{Methods}
\label{sec:methods}

\subsection{Non-adiabatic AHC theory}
\label{sec:ahc-theory}

The theory originally developed by AHC to describe the change of the electronic band structure due to electron-phonon interactions was derived from time-independent Rayleigh-Schrödinger perturbation theory.
In this case, the interaction between electrons and ions is described only statically, i.e., there is no energy transfer between electrons and phonons.
The approximation relies on the assumption that the phonon frequencies are much smaller than the typical electronic excitation energies.
Hence, one can assume that the electrons instantaneously adopt their electronic ground state for any ionic configuration.
This is the well known Born-Oppenheimer or adiabatic approximation.
The non-adiabatic case can be obtained rigorously using many-body perturbation theory treating electrons and ions on an equal quantum-mechanical footing.
This inherently allows for energy transfer from the electrons to the phonons and vice versa~\cite{giustino-review}.

\begin{figure}[!ht]
    \centering
    \includegraphics[width=0.48\textwidth]{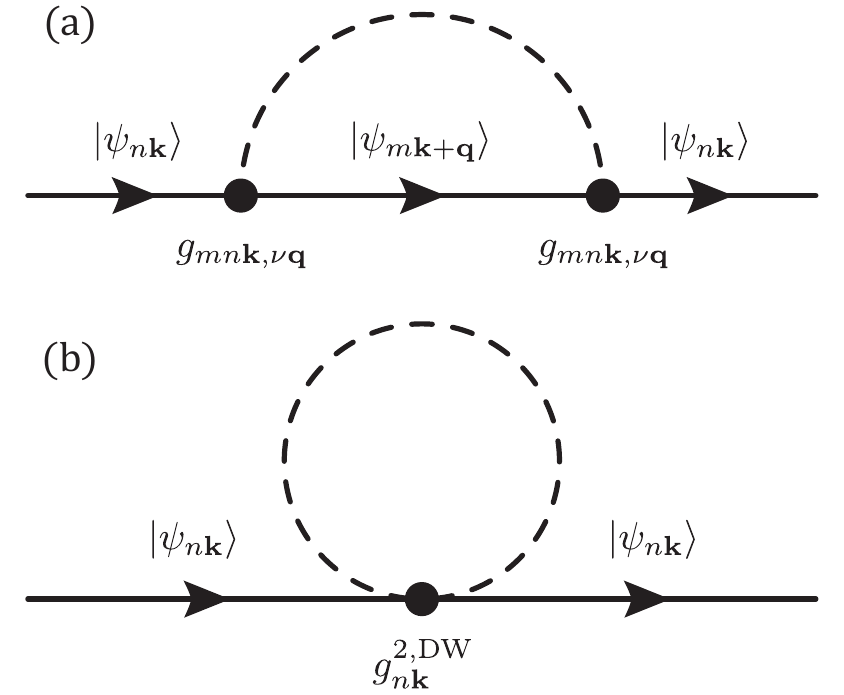}
    \caption{\label{fig:feynman}
        Feynman-diagrammatic representation of the lowest-order electron-phonon contributions to the electron self energy in the context of KS DFT.
        Electronic states (solid lines) interact with phonon states (dashed lines) via the first and second-order electron-phonon matrix elements, \(g_{mn \vec{k}, \nu \vec{q}} \) and \(g_\nk^{2,\text{DW}} \), respectively.
        (a) Fan-Migdal process; a phonon with wave vector \(\vec{q} \) and branch index \(\nu \) is emitted and later reabsorbed.
        (b) Debye-Waller process; a phonon is simultaneously emitted and reabsorbed.
        Note that there is no momentum transfer in this case.
    }
\end{figure}

If the perturbative expansion is truncated at second order in the phonon perturbation, one obtains two contributions to the change of the KS eigenvalues as a function of temperature, \(T \):
\begin{equation} \label{eq:renorm-split}
    \Delta \ev_\nk\paren{T} = \Delta \ev^\text{FM}_\nk\paren{T} + \Delta \ev^\text{DW}_\nk\paren{T}
    .
\end{equation}
The term \(\Delta \ev^\text{FM}_\nk\paren{T} \) is the real part of the so-called Fan-Migdal (FM) contribution to the electron self energy and corresponds to two first-order electron-phonon vertices.
\cref{fig:feynman}~(a) depicts the Feynman diagram associated with this process.
The FM contribution can be calculated as
\begin{multline} \label{eq:FM-temp}
	\Delta \ev^\text{FM}_\nk\paren{T}
	=
	\bzint \kint{q} \sum_\nu \sum_{m}
    \abs{g_{mn \vec{k}, \nu \vec{q}}}^2
    \\
	\times \mathrm{Re} \Bigg[
        \frac{1 - f_{m \vec{k} + \vec{q}}\paren{T} + n_{\nu \vec{q}}\paren{T}}{
            \ev_\nk - \ev_{m \vec{k} + \vec{q}} - \hbar \omega_{\nu \vec{q}} + \ci \delta
        }
        \\ +
        \frac{f_{m \vec{k} + \vec{q}}\paren{T} + n_{\nu \vec{q}}\paren{T}}{
            \ev_\nk - \ev_{m \vec{k} + \vec{q}} + \hbar \omega_{\nu \vec{q}} + \ci \delta
        }
    \Bigg]
	,
\end{multline}
where \(g_{mn \vec{k}, \nu \vec{q}} \) is the electron-phonon matrix element that describes scattering of an initial electronic state, \(\ket{\psi_\nk} \), into a final electronic state, \(\ket{\psi_{m \vec{k} + \vec{q}}} \), by either emitting or absorbing a phonon with wave vector \(\vec{q} \) and branch index \(\nu \).
The matrix element is discussed further in \cref{sec:elphon-paw} in relation to the PAW framework.
\(\ev_\nk \) are the electronic eigenvalues and \(\omega_{\nu \vec{q}} \) are the phonon angular frequencies.
\(f_{m \vec{k} + \vec{q}}\paren{T} \) and \(n_{\nu \vec{q}}\paren{T} \) are temperature-dependent Fermi-Dirac and Bose-Einstein distribution functions that correspond to the intermediate electronic and phononic states, respectively.
The integration boundary indicates that the domain of integration for the phonon wave vector is the first Brillouin zone with volume \(\BZvol \).
Finally, \(\delta \) is a positive infinitesimal that guarantees the correct pole structure of the self-energy in the complex plane.
From a practical perspective, this parameter can also be used to perform a Lorentzian broadening of the energy transitions.
This can improve convergence of the Brillouin-zone integration, but the parameter should be kept small to reduce potential numerical errors.
At zero temperature, \cref{eq:FM-temp} simplifies to the following expression:
\begin{equation} \label{eq:FM-zpr}
    \begin{split}
        \text{ZPR}^\text{FM}_\nk
        & \equiv \Delta \ev^\text{FM}_\nk\paren{T=0}
        \\ & = 
        \bzint \kint{q} \sum_\nu \sum_{m} \abs{g_{mn \vec{k}, \nu \vec{q}}}^2
        \\ & \times \mathrm{Re} \Bigg[
            \frac{1}{
                \ev_\nk - \ev_{m \vec{k} + \vec{q}} \pm \hbar \omega_{\nu \vec{q}} + \ci \delta
            }
        \Bigg]
        ,
    \end{split}
\end{equation}
where the sign in front of the phonon frequency in the denominator is positive if \(m \) corresponds to an occupied state and negative otherwise.

The remaining term in \cref{eq:renorm-split}, \(\Delta \ev^\text{DW}_\nk\paren{T} \), is the so-called Debye-Waller (DW) contribution and is described by only a single second-order perturbation.
Its associated Feynman diagram is shown in \cref{fig:feynman}~(b).
Unfortunately, the corresponding electron-phonon matrix elements, \(g_\nk^{2,\text{DW}} \), are expensive to calculate.
To resolve this issue, AHC introduced the rigid-ion approximation that allows the DW contribution to be written in terms of first-order electron-phonon matrix elements, \(g_{mn \vec{k}, \nu \vec{q}} \).
In this case, the DW contribution is approximated by
\begin{multline} \label{eq:DW-temp}
	\Delta \ev^\text{DW}_\nk\paren{T}
	= -
	\bzint \kint{q} \sum_\nu  \sum_{m}'
	\frac{\Xi_{mn\vec k, \nu \vec q}}{\ev_\nk - \ev_{m \vec k}}
    \\
	\times \paren{2 n_{\nu \vec q}\paren{T} + 1}
	,
\end{multline}
where
\begin{align}
	\Xi_{mn\vec k, \nu \vec q}
	& \equiv
	\frac{\hbar}{4 \omega_{\nu\vec q}}
	\sum_{\substack{\kappa\alpha \\ \kappa' \beta}}
	\Theta_{\kappa\alpha, \kappa'\beta}^{\nu \vec q}
	g^{0*}_{mn\vec k, \kappa \alpha}
	g^0_{mn\vec k, \kappa' \beta}
	, \\
	\Theta_{\kappa\alpha, \kappa'\beta}^{\nu \vec q}
	& \equiv
	\frac{e_{\kappa\alpha, \nu\vec q} e^*_{\kappa\beta, \nu\vec q}}{\mass_\kappa}
	+
	\frac{e^*_{\kappa'\alpha, \nu\vec q} e_{\kappa'\beta, \nu\vec q}}{\mass_{\kappa'}}
	, \\
	g^0_{mn\vec k, \kappa \alpha}
	& \equiv
	\sum_\nu
	\sqrt{\frac{2 \mass_\kappa \omega_{\nu\vec 0}}{\hbar}}
	e_{\kappa \alpha, \nu\vec 0} g_{mn\vec k, \nu\vec 0}
    .
\end{align}
The prime in the sum over all bands, \(m \), in \cref{eq:DW-temp} indicates that degenerate cases, where the denominator would become zero, are excluded.
The vector \(\vec{e}_{\kappa, \nu \vec{q}} \) describes the eigen-displacement of the atom \(\kappa \) with mass \(\mass_\kappa \) for the phonon mode characterized by \(\vec{q} \) and \(\nu \).
They are the eigenvectors of the dynamical matrix, \(D_{\kappa\alpha, \kappa'\beta}\paren{\vec{q}} \), which is defined as the Fourier transform of the matrix of inter-atomic force constants, \(C_{l\kappa\alpha, l'\kappa'\beta} \):
\begin{align}
    D_{\kappa\alpha, \kappa'\beta}\paren{\vec{q}}
    & \equiv
    \frac{1}{\sqrt{\mass_\kappa \mass_{\kappa'}}} \sum_l C_{l\kappa\alpha, 0\kappa'\beta} \eul^{-\ci \vec{q} \cdot \vec{R}_l}
    , \\
    C_{l\kappa\alpha, l'\kappa'\beta}
    & \equiv
    \frac{\partial^2 U}{\partial R_{l\kappa\alpha} \partial R_{l'\kappa'\beta}}
    .
\end{align}
The latter is defined as the second derivative of the total energy, \(U \), with respect to atomic displacements evaluated at the equilibrium configuration.
At zero temperature, we obtain the corresponding DW contribution to the ZPR:
\begin{equation} \label{eq:DW-zpr}
    \begin{split}
        \text{ZPR}^\text{DW}_\nk
        & \equiv
        \Delta \ev^\text{DW}_\nk \paren{T=0}
        \\ & = -
        \bzint \kint{q} \sum_\nu  \sum_{m}'
        \frac{\Xi_{mn\vec k, \nu \vec q}}{\ev_\nk - \ev_{m \vec k}}
        .
    \end{split}
\end{equation}

\subsection{Electron-phonon interactions within the PAW framework}
\label{sec:elphon-paw}

The pseudopotential method and related methods have been used for many decades to great success in solving the ground-state problem in DFT.
Ordinarily, when solving the KS equations, all electrons need to be considered.
This is problematic from a practical point of view.
Core electrons exhibit sharp features in real space, and due to the orthogonality constraint, the valence electrons also show rapid oscillations in real space close to the ionic cores.
In order to accurately capture these features using plane-wave basis sets, numerous plane-wave coefficients would be required.

Instead of treating all electrons explicitly, the pseudopotential method replaces the core electrons and nuclei by a soft, effective and singularity-free potential, a so-called pseudopotential.
This way, only a few valence states enter the KS equations, which allows for a considerable increase in computational efficiency.
In the original pseudopotential method, the exact orbitals are replaced by node-less pseudo (PS) orbitals and the shape of the AE orbitals cannot be easily recovered.

One particularly successful generalization of the pseudopotential method is the PAW method~\cite{bloechl-paw,kresse-paw}.
While this method retains some computational strategies from pseudopotential methods, the full-potential AE wave function can be reconstructed in the vicinity of each atom in the form of one-center terms that are represented on radial grids.
This allows for a very accurate description of both core and valence electrons while at the same time retaining many of the computational advantages of pseudopotential methods.
The price one has to pay is the emergence of additional terms in the expectation values of operators.
This is also very relevant for the definition of the electron-phonon matrix element within the PAW framework.

In terms of the AE orbitals, \(\ket{\psi_\nk} \), the electron-phonon matrix element can be computed as
\begin{align}
    \label{eq:g-ae}
    g_{mn \vec{k}, \nu \vec{q}}
    & \equiv
    \braket{\psi_{m \vec k + \vec q} | \partial_{\nu \vec q} \hat{H} | \psi_\nk}
    ,\\
    \partial_{\nu \vec q}
    & \label{eq:phonon-deriv}
    \equiv
    \sum_{l\kappa\alpha}
    \sqrt{\frac{\hbar}{2 \mass_\kappa \omega_{\nu \vec q}}}
    e_{\kappa\alpha, \nu \vec q}
    \eul^{\ci \vec q \cdot \vec R_{l}}
    \plka
    ,
\end{align}
where \(\hat{H} \) is the AE KS Hamiltonian and \(\partial_{\nu \vec q} \) corresponds to a collective displacement of the crystal lattice in terms of individual atomic displacements, \(\plka \equiv \frac{\partial}{\partial R_{l\kappa\alpha}} \).
In the PAW method, the AE orbitals are transformed into computationally convenient PS orbitals, \(\ket{\ppsi_\nk} \), via the PAW transformation:
\begin{equation}
    \ket{\ppsi_\nk} \equiv \pawT \ket{\psi_\nk}
    .
\end{equation}
Chaput~\etal{}~\cite{chaput-elphon} applied this transformation to express \(g_{mn \vec{k}, \nu \vec{q}} \) in \cref{eq:g-ae} in terms of the PAW quantities.
The corresponding final expression is given in \cref{app:ep-details} by \cref{eq:g}.
To obtain this result, the derivatives in \cref{eq:g-ae} are done at the level of the AE orbitals first, and then the PAW transformation is performed.
The electron-phonon matrix element so obtained is the one traditionally used in many-body perturbation theory.
We choose to call \(g_{mn \vec{k}, \nu \vec{q}} \) the AE electron-phonon matrix element.

Another strategy of taking the electron-phonon interaction into account was used by Engel~\etal{} in Ref.~\cite{engel-elph-paw}.
Indeed, in the adiabatic case, they showed that ZPR calculations within the PAW framework are also possible using an alternative definition of the electron-phonon matrix element,
\begin{equation} \label{eq:g-ps}
    \paw{g}_{mn \vec{k}, \nu \vec{q}}
    \equiv
    \braket{
        \ppsi_{m \vec k + \vec q} |
        \partial_{\nu \vec q} \paw{H} - \ev_\nk \partial_{\nu \vec q} \paw{S} |
        \ppsi_\nk
    }
    ,
\end{equation}
where \(\paw{H} = \pawT^\dagger \hat{H} \pawT \) is the PAW Hamiltonian and \(\paw{S} = \pawT^\dagger \pawT \) the PAW overlap operator.
To obtain this new quantity, the ZPR is expressed in terms of the derivatives of the electronic eigenvalues with respect to atomic displacement (Eq.~(2) in Ref.~\cite{engel-elph-paw}), and those electronic eigenvalues are obtained from the already pseudized Hamiltonian.
Casting the resulting equation for the ZPR into a form reminiscent of the one known from many-body perturbation theory allows to define \(\paw{g}_{mn \vec{k}, \nu \vec{q}} \), which we name the PS electron-phonon matrix element.
The matrix element so defined allows to rigorously compute the ZPR in the adiabatic approximation, however it is not hermitian.
Therefore it cannot be straightforwardly used within many-body perturbation theory and must be interpreted with care.

The key difference between these two approaches lies in the order in which the PAW transformation and the atomic derivatives are performed.
In the work of Engel~\etal{}~\cite{engel-elph-paw}, the quantity we want to differentiate (the electronic energies) is first expressed in terms of the PAW quantities, and then differentiated.
This is in the spirit of the original PAW formulation~\cite{bloechl-paw}: the total-energy expression is transformed using the PAW transformation and, subsequently, any other quantity is derived as a derivative thereof.
For instance, the Hamiltonian is the derivative of the PAW total energy with respect to the pseudo density matrix, the forces are the first derivative of the PAW total energy with respect to the ionic positions, inter-atomic force constants are the second derivative of the energy with respect to the ionic positions, and so on.
On the other hand, in their work Chaput~\etal{}~\cite{chaput-elphon} did not try to compute a specific physical quantity, like the ZPR.
Rather, they derived a computable expression, in terms of PAW quantities, for the full-potential AE electron-phonon matrix element, \(g_{mn \vec{k}, \nu \vec{q}} \).
Because it is defined from the derivative of the AE Hamiltonian, the atomic derivatives therefore have to be performed before the PAW transformation.

Remarkably, the two versions of the electron-phonon matrix element are related through the simple equation~\cite{engel-elph-paw}
\begin{align} \label{eq:g-ae-vs-ps}
    g_{mn \vec{k}, \nu \vec{q}}
    & =
    \paw{g}_{mn \vec{k}, \nu \vec{q}}
    +
    \paren{\ev_\nk - \ev_{m \vec{k} + \vec{q}}}
    t_{mn \vec{k}, \nu \vec{q}}
    , \\
    t_{mn \vec{k}, \nu \vec{q}}
    & \equiv
    \braket{
        \ppsi_{m \vec{k} + \vec{q}} |
        \pawT^{\dagger} \partial_{\nu \vec{q}} \pawT |
        \ppsi_\nk
    }
    .
\end{align}
Note that this involves derivatives of the PAW transformation operator, \(\pawT \), which are usually absent within the PAW framework, but emerge because the derivatives are taken first in Ref.~\cite{chaput-elphon}.
To compute the ZPR, both approaches are formally equivalent in the adiabatic case under the absence of the rigid-ion approximation (see \cref{app:ae-ps-equiv-proof}), but not necessarily in the non-adiabatic case or when the rigid-ion approximation is used.
Pseudized formulations of the electron-phonon matrix element have been used to calculate the non-adiabatic band-gap ZPR for many years, albeit most of the time using norm-conserving pseudopotentials~\cite{ponce-temperature-2014,ponce-static-vs-dynamic,nery-zpr-2018,giustino-diamond,cannuccia-effect-2011,cannuccia-zpr-2012,kawai-gan-2014,ponce-code-comparison-2014}.
In \cref{sec:results} we provide numerical evidence that using the PS electron-phonon matrix element is accurate to compute the ZPR, and even converges faster than the AE approach with respect to the number of intermediate states.
This can be rationalized by considering that \(\paw{g}_{mn \vec{k}, \nu \vec{q}} \) has been specifically designed to compute the ZPR in the adiabatic limit.
A more detailed explanation is given in \cref{sec:results}.

\subsection{Computational approach}
\label{sec:comp-approach}

The implementation used to calculate the ZPR in this work is based on the non-adiabatic AHC formulas in \cref{eq:FM-zpr,eq:DW-zpr} and utilizes finite atomic displacements in large supercells to evaluate the involved derivatives, \(\plka \), numerically.
Most of the code is implemented directly in VASP, but the derivatives are computed using a complementary python program.
The ZPR can both be evaluated using the PS and the AE electron-phonon matrix elements defined in \cref{sec:elphon-paw}.
In either case, a series of perturbations consisting of single atomic displacements, \(\plka \), are performed in large supercells.
From these calculations, both the inter-atomic force constants and the changes of the self-consistent KS potential are computed in the supercell and written to disk.
In VASP, the supercell potentials and force constants are read from disk, mapped to the primitive cell and evaluated between Bloch states corresponding to \(\vec{k} \) and \(\vec{k} + \vec{q} \) according to \cref{eq:g-ae}.
Note that VASP recalculates on-the-fly the Bloch orbitals on the required dense k-point grid.
In order to reconstruct the operator \(\partial_{\nu \vec{q}} \), linear combinations involving the phonon eigenvectors, \cref{eq:phonon-deriv}, are built.
This allows to compute \(\partial_{\nu \vec{q}} \paw{H} \), the derivative of the pseudopotential, \( \partial_{\nu \vec{q}} \paw{V} \), and other related quantities.
Details are given in \cref{app:ep-details} and in Ref.~\cite{chaput-elphon}.
The computation of the electron-phonon matrix elements using the derivative of the potential known in a supercell requires an interpolation procedure, which is detailed in \cref{app:interpolation}.
There, it is also shown that the procedure becomes exact when \(\vec{k} \) and \(\vec{q} \) are commensurate with the supercell used in the calculation.
For the related case of phonons, a derivation is also given in Ref.~\cite{parlinski-phonon}.

For ionic compounds, the long-range part of the potential derivatives contained in the electron-phonon matrix element must be treated explicitly.
To this end, a strategy similar to the one presented in Ref.~\cite{verdi-polar} is employed.
Details are given in \cref{app:long-range}.

Ultimately, the procedure outlined above allows to obtain both matrix elements, \(g_{mn \vec{k}, \nu \vec{q}} \) and \(\paw{g}_{mn \vec{k}, \nu \vec{q}} \), at arbitrary \(\vec{k} \) and \(\vec{q} \) vectors.
In order to estimate the integral over the first Brillouin zone in \cref{eq:FM-zpr,eq:DW-zpr}, an extrapolation towards an infinitely dense q-point grid is necessary.
It is assumed that at sufficiently dense q-point sampling densities, \(n_\text{q} \), the ZPR depends linearly on \(n_\text{q}^{-1/3} \), so that the final ZPR can be extrapolated from this linear regime~\cite{ponce-static-vs-dynamic}.

\section{Results}
\label{sec:results}

\subsection{AE versus PS formulation}
\label{sec:results-ae-vs-ps}

\begin{figure}[!ht]
    \centering
    \includegraphics[width=0.48\textwidth]{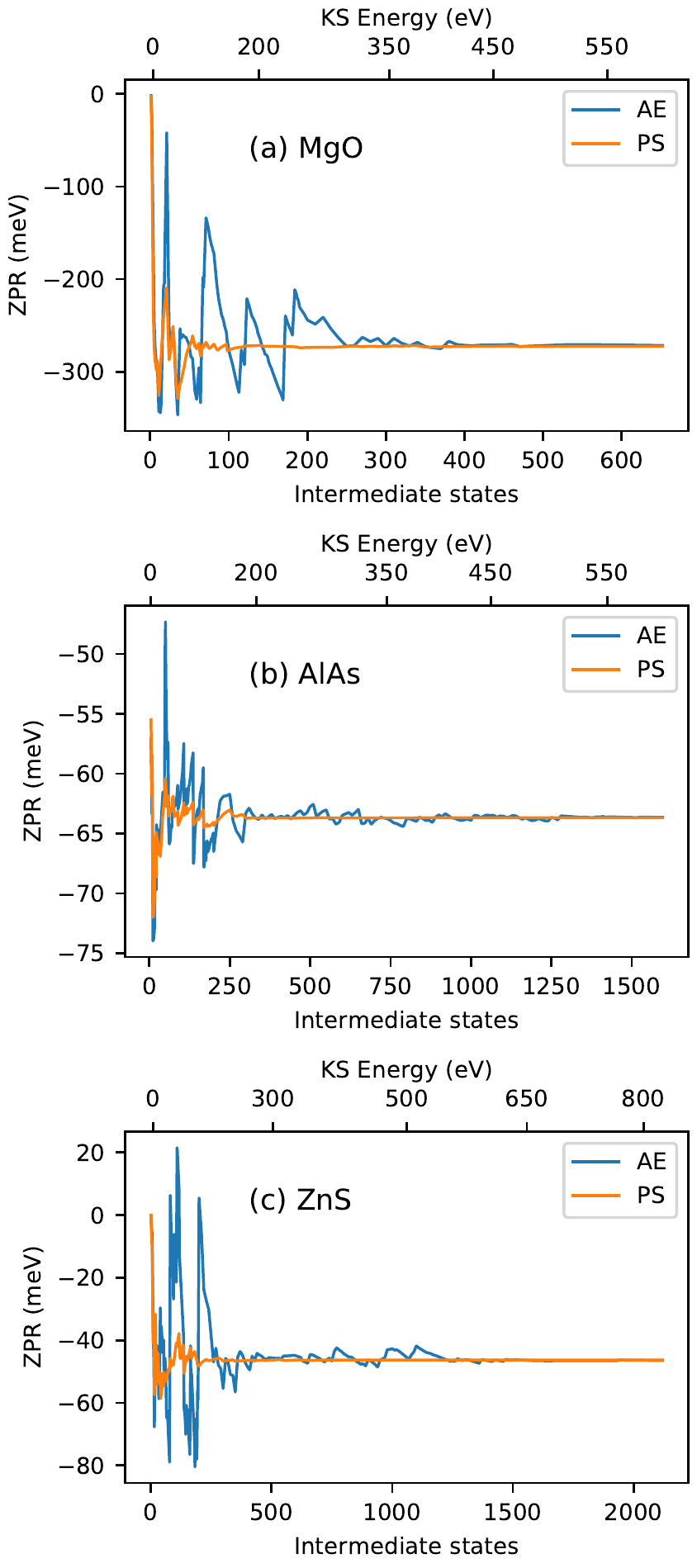}
    \caption{\label{fig:mats-ae-vs-ps}
        Convergence behavior of the non-adiabatic ZPR with respect to the maximum number of included intermediate states for both the AE and PS method for (a) MgO, (b) AlAs and (c) ZnS.
        The second x-axis scale above each panel displays the average KS energy with respect to the Fermi level that corresponds to the number of bands.
        The PS method converges much faster and smoother than the AE method, but both eventually converge to approximately the same value.
    }
\end{figure}

\begin{figure*}[!ht]
    \centering
    \includegraphics[width=0.98\textwidth]{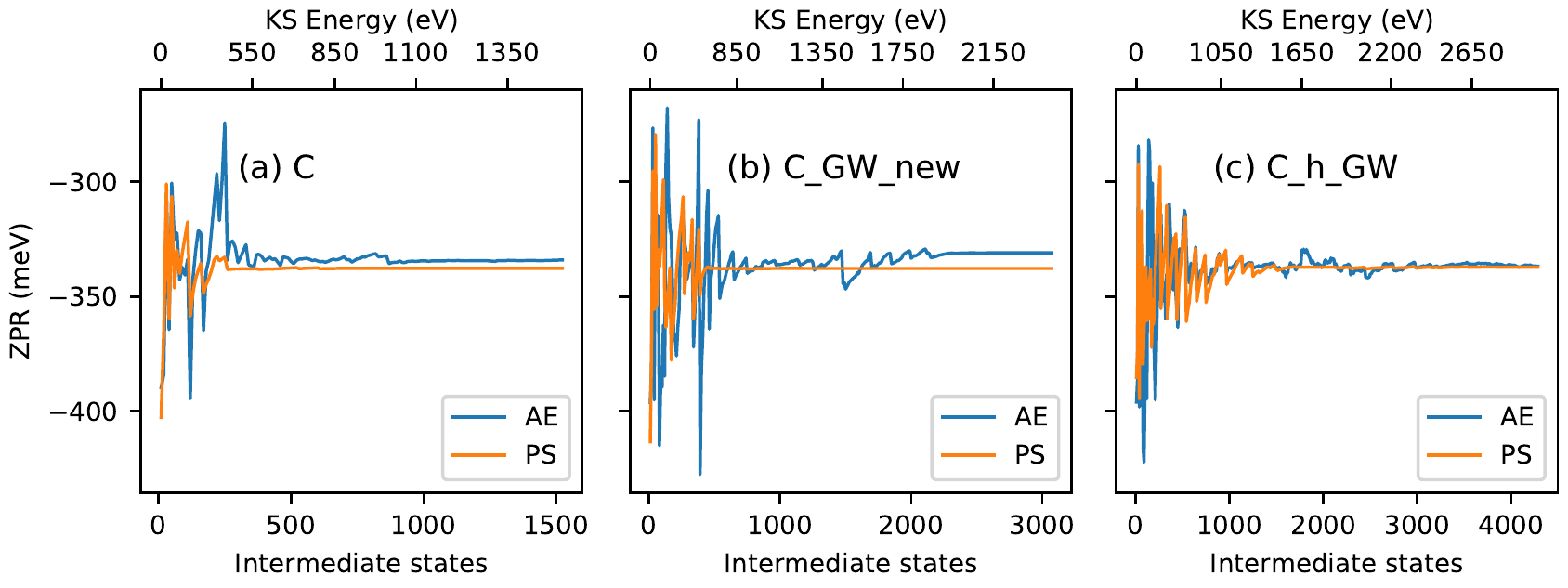}
    \caption{\label{fig:diamond-ae-vs-ps}
        Same as \cref{fig:mats-ae-vs-ps} but for diamond and using three different PAW potentials:
        C (a), C\_GW\_new (b) and C\_h\_GW (c).
        AE and PS methods converge to approximately the same result only in panel (c).
        The AE results depend more strongly on the PAW potential.
    }
\end{figure*}

First, we present a comparison between the AE and PS approaches discussed in \cref{sec:elphon-paw}.
To this end, we calculate the non-adiabatic band-gap ZPR for a few materials, namely MgO-rs, AlAs-zb, ZnS-zb and C-cd.
Here and in the following, the suffixes -rs, -zb, -cd and -w are used to denote the rock salt, zincblende, cubic diamond and wurtzite crystal structures, respectively.
For each of the four materials, two electron-phonon calculations are performed, one using the AE matrix elements and one using the PS matrix elements.
Both commence from the same \gridc{4} supercell calculation to determine the force-constants and the change of the self-consistent KS potential.
In our experience, the convergence of the ZPR with respect to the number of bands is largely independent of the convergence with respect to the number of q-points used to sample the phononic states.
Since we aim to highlight the difference between the AE and PS approach, we only use a relatively coarse \gridc{8} q-point mesh to sample the phononic states in each electron-phonon calculation.
This is identical to the \gridc{8} k-point mesh used to diagonalize the electronic KS Hamiltonian in the primitive cell.
The smearing parameter, \(\delta \), appearing in the denominator of \cref{eq:FM-zpr} is set to \SI{10}{\milli\electronvolt}.
The converged results for the non-adiabatic band-gap ZPR are listed in \cref{tab:ae-vs-ps} together with the PAW potentials used and the electronic cut-off energies.
We opt to use the labels of the POTCAR files distributed with VASP to distinguish between different PAW potentials.
Additional information that characterizes these PAW potentials is provided in \cref{tab:potcars}.
\begin{table}[!ht]
    \caption{\label{tab:ae-vs-ps}
        Comparison of the non-adiabatic band-gap ZPR converged with respect to the number of bands between AE and PS methods.
        In addition, the used PAW potentials (POTCAR) and plane-wave cutoffs (\(E_\text{cut} \)) are listed.
    }
    \begin{ruledtabular}
        \begin{tabular}{l r r r r}
            material &
            POTCAR &
            \(E_\text{cut} \) (\si{\electronvolt}) &
            ZPR AE &
            ZPR PS \\

            \colrule
            AlAs-zb & Al As         & 600   & -64   & -64   \\
            C-cd    & C             & 1500  & -334  & -338  \\
            C-cd    & C\_GW\_new    & 2400  & -331  & -338  \\
            C-cd    & C\_h\_GW      & 3000  & -337  & -337  \\
            MgO-rs  & Mg O\_s       & 600   & -272  & -273  \\
            ZnS-zb  & Zn S          & 800   & -46   & -46   \\
        \end{tabular}
    \end{ruledtabular}
\end{table}
The convergence behavior with respect to the number of bands is shown in \cref{fig:mats-ae-vs-ps,fig:diamond-ae-vs-ps}.
Each panel shows the non-adiabatic band-gap ZPR as a function of the maximum number of included intermediate states.
The secondary x-axis atop each panel displays the average KS energy associated with each intermediate state relative to the Fermi level.
Panels (a), (b) and (c) of \cref{fig:mats-ae-vs-ps} show the ZPR of MgO, AlAs and ZnS, respectively, while \cref{fig:diamond-ae-vs-ps} is dedicated to diamond.

It is known that the convergence of the ZPR with respect to the number of bands is slow and non-monotonic~\cite{giustino-diamond}, requiring many conduction-band states to reach convergence.
However, in \cref{fig:mats-ae-vs-ps,fig:diamond-ae-vs-ps}, we observe that the PS method converges appreciably faster than the AE method.
In order to understand why the ZPR converges slowly in the first place, let us consider the FM contribution in the adiabatic approximation.
The relevant quantity is the integrand \(\gamma^{\text{FM}}_{\nk,\nu \vec{q}} \) in \cref{eq:app:zpr-fm,eq:app:zpr-fm-gamma-ae}.
When written using the one-particle Green's function, \(\paren{\ev_\nk - \hat{H}}^{-1} \), the integrand becomes:
\begin{multline}
    \gamma^{\text{FM}}_{\nk,\nu \vec{q}}
    = \sum_{m}'
    \braket{\psi_\nk | \partial_{\nu -\vec q} \hat{H} | \psi_{m \vec k + \vec q}}
    \\ \times
    \braket{\psi_{m \vec k + \vec q} | \paren{\ev_\nk -  \hat{H}}^{-1} \partial_{\nu \vec q} \hat{H} | \psi_\nk}
    .
\end{multline}
In this form, it becomes clear that the sum over all intermediate states, \(\ket{\psi_{m \vec{k} + \vec{q}}} \), can be considered an expansion of the perturbed AE potential present in  \( \partial_{\nu \vec q} \hat{H} \) in terms of KS orbitals.
Since the KS orbitals at high energies essentially behave like free electrons, the sum eventually becomes a plane-wave expansion at high energies.
Unfortunately, the true AE potential and its derivative can change rapidly in proximity to the nuclei.
Thus, in general, a considerable amount of plane waves is required to accurately represent the change of the AE potential.
This explains the origin of the slow convergence of the ZPR with respect to the number of intermediate states.
Similar arguments can be made to discuss the convergence of the DW part of the ZPR.

Let us now attempt to explain the difference in convergence behavior between the AE and PS approaches as seen in \cref{fig:mats-ae-vs-ps,fig:diamond-ae-vs-ps}.
In the PS approach, the PS FM term is calculated by replacing the integrand \(\gamma^{\text{FM}}_{\nk,\nu \vec{q}} \) in \cref{eq:app:zpr-fm} by its PS counterpart, \(\paw{\gamma}^{\text{FM}}_{\nk,\nu \vec{q}} \), defined in \cref{eq:app:zpr-fm-gamma-ps}.
The two integrands differ only by a substitution \(g_{mn \vec{k}, \nu \vec{q}} \to \paw{g}_{mn \vec{k}, \nu \vec{q}} \) and by the last term in \cref{eq:app:zpr-fm-gamma-ps}.
This term has been carefully checked to be negligible for all materials considered here.
Therefore, the difference in convergence behavior between the AE and PS approaches is directly related to the difference between the corresponding electron-phonon matrix elements.
As a side note, the individual FM and DW terms generally converge to completely different values in the two formulations, as discussed in Ref.~\cite{engel-elph-paw}.

In the PAW method, slow plane-wave convergence is usually avoided since the sharp AE potential is replaced by a smooth pseudopotential that requires fewer plane waves to be described accurately.
In the case of the PS electron-phonon matrix element, this pseudization is kept intact as we pseudize the orbitals and potentials first, and only describe the interaction of the PS orbitals with the pseudopotentials and their derivatives.
Although the pseudized electron-phonon term, \cref{eq:g-ps}, contains one-center corrections --- as happens for any operator after the PAW transformation --- these one-center terms are calculated in such a manner that the ionic potential and partial waves move rigidly in unison with the PAW sphere when the corresponding atom is displaced.
As a consequence, there are no partial derivatives of the AE potential.
This explains the comparatively faster convergence in the PS case.

On the other hand, in the case of the AE electron-phonon matrix element, the change of the AE potential is essentially reconstructed from the available information inside the PAW spheres using \cref{eq:g,eq:gv,eq:gd,eq:gp,eq:gr}.
This reintroduces the slow convergence as discussed earlier.
To see how this manifests in practice, let us inspect the difference between the AE and PS matrix elements given in \cref{eq:g-ae-vs-ps}.
It factorizes into two parts: an energy difference, \(\paren{\ev_\nk - \ev_{m \vec{k} + \vec{q}}} \), and the term \(t_{mn \vec{k}, \nu \vec{q}} \) that contains the change of the AE partial waves (see \cref{eq:gr}).
The energy difference causes this contribution to converge slowly as it scales linearly with the KS energy of the intermediate state, while \(t_{mn \vec{k}, \nu \vec{q}} \) provides a numerical reconstruction of the perturbed AE potential.
In contrast, no such scaling is present in the PS method as the corresponding matrix elements do not contain the KS energy of the intermediate state.

If one were to artificially set \(\paren{\ev_\nk - \ev_{m \vec{k} + \vec{q}}} = 1 \) in \cref{eq:gp,eq:gr}, the fast convergence behavior of the PS method should be recovered.
We have conducted this numerical test for the carbon potential corresponding to panel (a) of \cref{fig:diamond-ae-vs-ps} and indeed found that the modified ZPR calculation converges as fast as in the PS case.
Once again, this is not a coincidence as the reconstruction of the AE information contained in \(t_{mn \vec{k}, \nu \vec{q}} \) is not required in the PS method.

For diamond, as observed in panels (a) and (b) of \cref{fig:diamond-ae-vs-ps}, both methods converge to slightly different values.
In our implementation, \cref{eq:g} can be seen as a reconstruction formula to obtain the AE quantity, \cref{eq:g-ae}.
That both methods do not converge to the same value indicates that this reconstruction in the term \(t_{mn \vec{k}, \nu \vec{q}} \) has failed.
More precisely, it is most likely the contribution \(\mathbf{g}^{(R)} \) in \cref{eq:gr}, which contains the derivative of the AE partial waves, \(\ket{\phi_{j}} \), projected onto other partial waves, \(\bra{\phi_{i}} \), that is responsible for the discrepancy.
Unfortunately, it is entirely possible that the corresponding partial-wave basis is not sufficiently complete to describe the potential derivative induced by the phonon perturbation, especially at high energies.
In general, pseudopotential methods, the PAW method and other linearized methods have been designed to describe total energies, forces, and one-electrons energies close to the Fermi-level with great accuracy~\cite{andersen-linear-1975}.
In contrast, electron-phonon matrix elements are, strictly speaking, not observables.
Hence, it is reasonable to assume that the reconstruction of the AE orbital might fail for certain pseudopotentials, especially at higher energies where the partial-wave basis is not sufficiently complete.
An important exception are the diagonal matrix elements (\(\ev_{m \vec{k} + \vec{q}} = \ev_\nk \)) which describe the first-order change of the electronic eigenvalues.
In this case, the AE and PS electron-phonon matrix elements are strictly identical and the contribution from \cref{eq:gr} is zero.
Similar issues related to the completeness of the local PAW basis were previously encountered in the context of linear optical properties~\cite{gajdos-lepsilon} and are also commonly observed for NMR calculations in all-electron codes~\cite{de-wijs-nmr-2017}.

In panel (b) of \cref{fig:diamond-ae-vs-ps}, we have tried to improve upon the calculations performed in panel (a) by increasing the number of partial-waves, but this did not improve the agreement.
In panel (c), we used a pseudopotential with a smaller core radius for the PAW spheres, a local pseudopotential that follows the all-electron potential down to 0.8 Bohr radii, and a partial-wave basis that contains three partial waves and accurately restores the scattering properties up to \SI{500}{\electronvolt} above the vacuum level.
This effectively shifts a portion of the AE information from the local basis onto the plane-wave grid and improves the completeness of the partial wave basis.
The drawback of such a potential is that an even larger number of states is required to converge the ZPR with respect to the number of intermediate states.
In panel (c), the AE and PS methods agree on the final result, as reported in \cref{tab:ae-vs-ps}.
Remarkably, the values computed using the PS approach seem to be quite insensitive to the choice of the pseudopotential, which motivates our choice to calculate the ZPR using the PS approach.

\subsection{Evaluation of ZPR for selected materials}
\label{sec:results-materials}

In light of the above results and discussion, from here on we have calculated the ZPR using the PS approach.
We now proceed to present converged non-adiabatic ZPR calculations for several materials listed in \cref{tab:pots}.
Important computational parameters, such as the choice of the PAW potential and the plane-wave cutoff, are reported.
\begin{table}[!ht]
    \caption{\label{tab:pots}
        PAW potentials (POTCAR), plane-wave cutoffs (\(E_\text{cut} \)) and supercell sizes used for all materials studied.
        In addition, the finest k/q-point grid size used to extrapolate the ZPR towards infinity is listed for each material (max. grid).
        This grid size is used for both electrons and phonons.
        The names of the PAW potentials correspond to the labels of the POTCAR files distributed with VASP.
    }
    \begin{ruledtabular}
        \begin{tabular}{l r c c c}
            material &
            POTCAR &
            \(E_\text{cut} \) (\si{\electronvolt}) &
            supercell &
            max. grid \\

            \colrule
            AlAs-zb & Al As       & 240 & \gridc{4}      & \gridc{64} \\
            AlN-w   & Al N\_s     & 279 & \grid{4}{4}{2} & \gridc{48} \\
            AlP-zb  & Al P        & 255 & \gridc{4}      & \gridc{64} \\
            AlSb-zb & Al Sb       & 240 & \gridc{4}      & \gridc{64} \\
            BN-zb   & B N\_s      & 319 & \gridc{4}      & \gridc{64} \\
            BaO-rs  & Ba\_sv O\_s & 283 & \gridc{4}      & \gridc{64} \\
            BeO-w   & Be O\_s     & 350 & \grid{4}{4}{2} & \gridc{48} \\
            C-cd    & C           & 600 & \gridc{4}      & \gridc{48} \\
            CaO-rs  & Ca\_sv O\_s & 350 & \gridc{4}      & \gridc{64} \\
            CdS-zb  & Cd S        & 274 & \gridc{4}      & \gridc{64} \\
            CdSe-zb & Cd Se       & 274 & \gridc{4}      & \gridc{64} \\
            CdTe-zb & Cd Te       & 274 & \gridc{4}      & \gridc{64} \\
            GaN-w   & Ga\_d N\_s  & 283 & \grid{4}{4}{2} & \gridc{64} \\
            GaN-zb  & Ga\_d N\_s  & 350 & \gridc{4}      & \gridc{64} \\
            GaP-zb  & Ga\_d P     & 283 & \gridc{4}      & \gridc{64} \\
            Li2O    & Li\_sv O    & 499 & \gridc{4}      & \gridc{64} \\
            LiF-rs  & Li\_sv F    & 499 & \gridc{4}      & \gridc{64} \\
            MgO-rs  & Mg O        & 500 & \gridc{4}      & \gridc{64} \\
            Si-cd   & Si          & 245 & \gridc{4}      & \gridc{64} \\
            SiC-zb  & Si C        & 400 & \gridc{4}      & \gridc{64} \\
            SiO2-t  & Si O\_s     & 283 & \grid{3}{3}{4} & \gridc{48} \\
            SnO2-t  & Sn\_d O\_s  & 283 & \grid{3}{3}{4} & \gridc{48} \\
            SrO-rs  & Sr\_sv O\_s & 283 & \gridc{4}      & \gridc{64} \\
            TiO2-t  & Ti\_sv O\_s & 283 & \grid{3}{3}{4} & \gridc{32} \\
            ZnO-w   & Zn O\_s     & 283 & \grid{4}{4}{2} & \gridc{64} \\
            ZnS-zb  & Zn S        & 277 & \gridc{4}      & \gridc{64} \\
            ZnSe-zb & Zn Se       & 277 & \gridc{4}      & \gridc{64} \\
            ZnTe-zb & Zn Te       & 400 & \gridc{4}      & \gridc{64} \\
        \end{tabular}
    \end{ruledtabular}
\end{table}
In addition, the table also lists the size of the supercell and the maximum number of k/q-points used to extrapolate the ZPR towards infinity.
Additional information regarding the pseudopotentials is listed in \cref{tab:potcars}.
The materials were chosen from a recent publication by Miglio~\etal{}~\cite{miglio-zpr} in order to make a detailed comparison.

All calculations were performed using the PBE flavor of density-functional-theory approximation~\cite{perdew-pbe-1,perdew-pbe-2}, except AlN, BN and C, where LDA was used~\cite{ceperley-lda-ca}.
This is done for consistency with the calculations reported by Miglio~\etal{}, as different density-functional-theory approximations can translate into slight differences in the ZPR results.
For reason of consistency, we also chose a smearing parameter of \(\delta = \SI{10}{\milli\electronvolt}\) for all compounds and the same band-gap transitions as in Ref.~\cite{miglio-zpr}.
Lattice parameters are optimized to minimize the DFT total-energy functional.
The Born effective-charge tensor and macroscopic ion-clamped static dielectric tensor used to calculate the long-range electrostatic contributions in our interpolation scheme are calculated from DFPT~\cite{gajdos-lepsilon}.
We leave the special treatment of the dynamic quadrupoles in the long-range interaction as proposed in~\cite{brunin-quadrupole,jhalani-quadrupole} for a future work.
Lattice parameters as well as the relevant components of the Born effective-charge and dielectric tensor are reported in \cref{tab:latt-born-cubic,tab:latt-born-hexagonal,tab:latt-born-tetragonal}.
\begin{table}[!ht]
    \caption{\label{tab:latt-born-cubic}
        Lattice constants (in \si{\angstrom}), dielectric constants and Born effective charges for cubic materials.
    }
    \begin{ruledtabular}
        \begin{tabular}{l c c c}
            material &
            \(a \) &
            \(\epsilon^\infty_{xx} \) &
            \(Z^\star_{xx} \) \\
            
            \colrule
            AlAs-zb &  4.053 &   9.526 &   2.154 \\
            AlP-zb  &  3.893 &   8.118 &   2.245 \\
            AlSb-zb &  4.407 &  12.035 &   1.827 \\
            BN-zb   &  2.535 &   4.534 &   1.866 \\
            BaO-rs  &  3.968 &   4.238 &   2.722 \\
            CaO-rs  &  3.421 &   3.768 &   2.346 \\
            CdS-zb  &  4.190 &   6.227 &   2.235 \\
            CdSe-zb &  4.379 &   8.386 &   2.319 \\
            CdTe-zb &  4.680 &   9.304 &   2.302 \\
            GaN-zb  &  3.215 &   6.059 &   2.669 \\
            GaP-zb  &  3.891 &  10.583 &   2.206 \\
            Li2O    &  3.268 &   2.918 &   0.903 \\
            LiF-rs  &  2.867 &   2.049 &   1.052 \\
            MgO-rs  &  2.978 &   3.128 &   1.969 \\
            SiC-zb  &  3.096 &   6.993 &   2.690 \\
            SrO-rs  &  3.682 &   3.781 &   2.431 \\
            ZnS-zb  &  3.851 &   5.905 &   2.022 \\
            ZnSe-zb &  4.055 &   7.345 &   2.102 \\
            ZnTe-zb &  4.372 &   9.030 &   2.087 \\
        \end{tabular}
    \end{ruledtabular}
\end{table}
\begin{table}[!ht]
    \caption{\label{tab:latt-born-hexagonal}
        Lattice constants (in \si{\angstrom}) and components of the dielectric tensor and Born effective-charge tensor for symmetry-inequivalent directions for hexagonal materials.
    }
    \begin{ruledtabular}
        \begin{tabular}{l c c c c c c}
            material &
            \(a \) & \(c \) &
            \(\epsilon^\infty_{xx} \) & \(\epsilon^\infty_{zz} \) &
            \(Z^\star_{xx} \) & \(Z^\star_{zz} \) \\
            
            \colrule
            AlN-w &  3.091 &  4.947 &   4.372 &   4.592 &   2.504 &   2.665 \\
            BeO-w &  2.712 &  4.404 &   3.044 &   3.108 &   1.788 &   1.848 \\
            GaN-w &  3.219 &  5.244 &   5.855 &   6.035 &   2.621 &   2.761 \\
            ZnO-w &  3.284 &  5.301 &   5.242 &   5.227 &   2.109 &   2.162 \\
        \end{tabular}
    \end{ruledtabular}
\end{table}
\begin{table}[!ht]
    \caption{\label{tab:latt-born-tetragonal}
        Same as \cref{tab:latt-born-hexagonal} but for tetragonal materials.
    }
    \begin{ruledtabular}
        \begin{tabular}{l c c c c c c c}
            material &
            \(a \) & \(c \) &
            \(\epsilon^\infty_{xx} \) & \(\epsilon^\infty_{zz} \) &
            \(Z^\star_{xx} \) & \(Z^\star_{xy} \) & \(Z^\star_{zz} \) \\
            
            \colrule
            SiO2-t &  4.231 &  2.700 &   3.369 &   3.556 &   3.807 &   0.401 &   4.041 \\
            SnO2-t &  4.831 &  3.246 &   4.675 &   4.909 &   4.097 &   0.519 &   4.474 \\
            TiO2-t &  4.587 &  2.954 &   7.404 &   8.738 &   6.341 &   1.030 &   7.621 \\
        \end{tabular}
    \end{ruledtabular}
\end{table}

\begin{table*}[!ht]
    \caption{\label{tab:potcars}
        List of PAW potentials used in this work.
        The first column (POTCAR) contains the label of the potential from VASP's PAW database.
        The second column (valence) shows which local orbitals are treated as valence.
        The cutoff radii, \(r_{\text{cut}} \), for the s, p, d and f channels are reported in columns three to six if applicable.
        If there is more than one channel for a single angular-momentum quantum number, the corresponding cutoff radii are listed as \(n \times r_{\text{cut}} \) if they are the same or as a comma-separated list otherwise.
        The final column contains the radius \(r_{\text{core}} \) below which the AE potential is replaced by a local pseudopotential.
        All radii are given in atomic units.
    }
    \begin{ruledtabular}
        \begin{tabular}{l c c c c c l}
            POTCAR &
            valence &
            \(r_{\text{cut}}(s) \) &
            \(r_{\text{cut}}(p) \) &
            \(r_{\text{cut}}(d) \) &
            \(r_{\text{cut}}(f) \) &
            \(r_{\text{core}} \) \\
            
            \colrule
            Al         & 3s3p3d     & \ensuremath{2 \times 1.9}  & \ensuremath{2 \times 1.9}  & 1.9                       & --  & 1.900 \\
            As         & 4s4p4d4f   & \ensuremath{2 \times 2.1}  & \ensuremath{2 \times 2.1}  & 2.1                       & 2.1 & 2.100 \\
            B          & 2s2p3d     & \ensuremath{2 \times 1.5}  & \ensuremath{2 \times 1.7}  & 1.7                       & --  & 1.700 \\
            Ba\_sv     & 5s6s5p5d   & \ensuremath{2 \times 2.8}  & \ensuremath{2 \times 2.7}  & \ensuremath{2 \times 2.7} & --  & 2.516 \\
            Be         & 2s2p3d     & \ensuremath{2 \times 1.9}  & \ensuremath{2 \times 1.9}  & 1.5                       & --  & 1.900 \\
            C          & 2s2p3d     & \ensuremath{2 \times 1.2}  & \ensuremath{2 \times 1.5}  & 1.5                       & --  & 1.500 \\
            C\_GW\_new & 2s2p       & \ensuremath{3 \times 1.1}  & \ensuremath{3 \times 1.5}  & 1.5, 1.6                  & 1.4 & 1.600 \\
            C\_h\_GW   & 2s2p       & \ensuremath{3 \times 1.0}  & \ensuremath{3 \times 1.1}  & \ensuremath{2 \times 1.1} & --  & 0.804 \\
            Ca\_sv     & 3s4s3p3d   & \ensuremath{2 \times 2.3}  & \ensuremath{2 \times 2.3}  & \ensuremath{2 \times 2.3} & --  & 1.808 \\
            Cd         & 4d5s       & \ensuremath{2 \times 2.3}  & \ensuremath{2 \times 2.3}  & \ensuremath{2 \times 2.3} & --  & 2.054 \\
            F          & 2s2p3d     & \ensuremath{2 \times 1.2}  & \ensuremath{2 \times 1.52} & 1.5                       & --  & 1.520 \\
            Ga\_d      & 3d4s4p     & \ensuremath{2 \times 2.3}  & 2.1, 2.3                   & \ensuremath{2 \times 2.3} & 2.3 & 2.300 \\
            Li\_sv     & 1s2s2p3d   & 1.4, 1.7                   & 1.4                        & 1.4                       & --  & 1.700 \\
            Mg         & 3s3d       & \ensuremath{2 \times 2.0}  & \ensuremath{2 \times 2.0}  & 2.0                       & --  & 1.506 \\
            N\_s       & 2s2p       & \ensuremath{2 \times 1.5}  & \ensuremath{2 \times 1.85} & --                        & --  & 0.803 \\
            O          & 2s2p3d     & \ensuremath{2 \times 1.2}  & \ensuremath{2 \times 1.52} & 1.5                       & --  & 1.520 \\
            O\_s       & 2s2p       & \ensuremath{2 \times 1.5}  & \ensuremath{2 \times 1.85} & --                        & --  & 0.804 \\
            P          & 3s3p3d     & \ensuremath{2 \times 1.9}  & \ensuremath{2 \times 1.9}  & 1.9                       & --  & 1.900 \\
            S          & 3s3p3d     & \ensuremath{2 \times 1.9}  & \ensuremath{2 \times 1.9}  & 1.9                       & --  & 1.900 \\
            Sb         & 5s5p5d4f   & \ensuremath{2 \times 2.3}  & \ensuremath{2 \times 2.3}  & 2.3                       & 2.3 & 2.300 \\
            Se         & 4s4p4d4f   & \ensuremath{2 \times 2.1}  & \ensuremath{2 \times 2.1}  & 2.1                       & 2.1 & 2.100 \\
            Si         & 3s3p3d     & \ensuremath{2 \times 1.9}  & \ensuremath{2 \times 1.9}  & 1.9                       & --  & 1.900 \\
            Sn\_d      & 4d5s5p     & \ensuremath{2 \times 2.5}  & \ensuremath{2 \times 2.5}  & \ensuremath{2 \times 2.5} & 2.5 & 2.500 \\
            Sr\_sv     & 4s5s4p4d   & \ensuremath{2 \times 2.48} & \ensuremath{2 \times 2.5}  & \ensuremath{2 \times 2.5} & --  & 2.201 \\
            Te         & 5s5p5d4f   & \ensuremath{2 \times 2.3}  & \ensuremath{2 \times 2.3}  & 2.3                       & 2.3 & 2.300 \\
            Ti\_sv     & 3s4s3p3d4f & 1.8, 2.3                   & \ensuremath{2 \times 2.3}  & \ensuremath{2 \times 2.3} & 2.3 & 2.300 \\
            Zn         & 3d4s       & \ensuremath{2 \times 2.3}  & \ensuremath{2 \times 2.3}  & \ensuremath{2 \times 2.3} & --  & 1.828 \\
        \end{tabular}
    \end{ruledtabular}
\end{table*}

Our results for the band-gap ZPR are reported in \cref{tab:zpr}, together with the corresponding results from Miglio~\etalNoPeriod{}.
\begin{table}[!ht]
    \centering
    \caption{\label{tab:zpr}
        Band-gap ZPR (in \si{\milli\electronvolt}) obtained in the framework of non-adiabatic AHC theory for various materials.
        The values are compared against the ones reported by Miglio~\etal{}~\cite{miglio-zpr}.
    }
    \begin{ruledtabular}
        \begin{tabular}{l r r r r}
            \multirow{2}*{material} &
            ZPR &
            ZPR &
            \multirow{2}*{rel.~diff.} &
            \multirow{2}*{abs.~diff.} \\
            
            &
            this work &
            Ref.~\cite{miglio-zpr} & \\

            \colrule
            AlAs-zb &   -74 &   -74 & \SI{0.2}{\percent}  &     0 \\
            AlN-w   &  -377 &  -399 & \SI{5.5}{\percent}  &    22 \\
            AlP-zb  &   -96 &   -93 & \SI{2.9}{\percent}  &     3 \\
            AlSb-zb &   -52 &   -51 & \SI{1.8}{\percent}  &     1 \\
            BN-zb   &  -402 &  -406 & \SI{0.9}{\percent}  &     4 \\
            BaO-rs  &  -277 &  -271 & \SI{2.4}{\percent}  &     6 \\
            BeO-w   &  -726 &  -699 & \SI{3.9}{\percent}  &    27 \\
            C-cd    &  -323 &  -330 & \SI{2.0}{\percent}  &     7 \\
            CaO-rs  &  -357 &  -341 & \SI{4.8}{\percent}  &    16 \\
            CdS-zb  &   -67 &   -70 & \SI{3.8}{\percent}  &     3 \\
            CdSe-zb &   -29 &   -34 & \SI{15.2}{\percent} &     5 \\
            CdTe-zb &   -19 &   -20 & \SI{7.4}{\percent}  &     1 \\
            GaN-w   &  -171 &  -189 & \SI{9.5}{\percent}  &    18 \\
            GaN-zb  &  -163 &  -176 & \SI{7.6}{\percent}  &    13 \\
            GaP-zb  &   -69 &   -65 & \SI{5.7}{\percent}  &     4 \\
            Li2O    &  -569 &  -573 & \SI{0.6}{\percent}  &     4 \\
            LiF-rs  & -1231 & --    & --                  & --    \\
            MgO-rs  &  -533 &  -524 & \SI{1.6}{\percent}  &     9 \\
            Si-cd   &   -58 &   -56 & \SI{4.1}{\percent}  &     2 \\
            SiC-zb  &  -175 &  -179 & \SI{2.1}{\percent}  &     4 \\
            SiO2-t  &  -583 &  -585 & \SI{0.4}{\percent}  &     2 \\
            SnO2-t  &  -232 &  -215 & \SI{7.9}{\percent}  &    17 \\
            SrO-rs  &  -323 &  -326 & \SI{1.0}{\percent}  &     3 \\
            TiO2-t  &  -349 &  -337 & \SI{3.5}{\percent}  &    12 \\
            ZnO-w   &  -175 &  -157 & \SI{11.2}{\percent} &    18 \\
            ZnS-zb  &   -88 &   -88 & \SI{0.2}{\percent}  &     0 \\
            ZnSe-zb &   -43 &   -44 & \SI{3.1}{\percent}  &     1 \\
            ZnTe-zb &   -25 &   -22 & \SI{14.4}{\percent} &     3 \\
        \end{tabular}
    \end{ruledtabular}
\end{table}
Given the differences in the pseudopotentials and the different methodological details, the comparison between the ZPR results is excellent.
For most materials, the relative difference is only within a few percent, except for some compounds that feature a particularly small ZPR, such as ZnTe and CdSe.
Concerning LiF, missing in Ref.~\cite{miglio-zpr}, we note that Nery~\etal{}~\cite{nery-zpr-2018} report a non-adiabatic band-gap ZPR for LiF of \SI{-1149}{\milli\electronvolt}.
This agrees well with our value of \SI{-1231}{\milli\electronvolt}.

In order to reaffirm the correctness of our results, we conducted a more thorough analysis on ZnO-w for which the relative and absolute differences are comparatively large.
First, we compare our computational setup with the one used by Miglio~\etal{} in Ref.~\cite{miglio-zpr}.
While small differences exist in the lattice parameters, Born effective charges and the dielectric tensor, we find that those differences only account for a change in the ZPR of about \SI{3}{\milli\electronvolt}.
Likewise, increasing the supercell size in our calculation from \grid{4}{4}{2} to \grid{5}{5}{3} increases the total ZPR by only \SI{2}{\milli\electronvolt}.

Next, we investigate the effect of using different pseudopotentials on the ZPR.
The AHC band-gap ZPR of ZnO reported in Ref.~\cite{miglio-zpr} was calculated in Abinit~\cite{gonze-abinit-2020} using ONCVPSP-type pseudopotentials from the Pseudo-Dojo database v0.3~\cite{van-setten-pseudo-dojo}.
Here, we attempt to replicate the result of Ref.~\cite{miglio-zpr}, and we perform additional calculations using the EPH module in Abinit 9.6.2 which uses a similar approach to interpolate the electron-phonon potential to dense q-point meshes.
We use the Dojo pseudopotentials in their standard and high variants, as well as two other types denoted FHI98pp and HGH following the denominations in Ref.~\cite{lejaeghere-reproducibility-2016}.

The results of our Abinit calculations are summarized in \cref{tab:zno}.
\begin{table}[!ht]
    \centering
    \caption{\label{tab:zno}
        Comparison between different norm-conserving pseudopotentials used in our Abinit calculations of the band-gab ZPR of ZnO-w (in \si{\milli\electronvolt}).
        The row labeled ``infinity'' contains the extrapolated values.
    }
    \begin{ruledtabular}
        \begin{tabular}{l r r r r}
            dense grid &
            FHI98pp &
            Dojo-std. &
            Dojo-high &
            HGH \\
            
            \colrule
            \gridc{4}   &  -81  &  -88  &  -88  &  -84  \\
            \gridc{8}   & -106  & -116  & -116  & -110  \\
            \gridc{12}  & -117  & -129  & -129  & -122  \\
            \gridc{16}  & -124  & -137  & -137  & -129  \\
            \gridc{24}  & -133  & -148  & -148  & -139  \\
            \gridc{32}  & -139  & -153  & -154  & -146  \\
            \gridc{48}  & -145  & -162  & -162  & -152  \\
            \gridc{64}  & -149  & -166  & -166  & -158  \\
            infinity    & -160  & -175  & -178  & -175  \\
        \end{tabular}
    \end{ruledtabular}
\end{table}
The HGH as well as both Dojo calculations yield very similar values when extrapolated to an infinitely dense q-point grid, in excellent agreement with the ones obtained by VASP using the PAW method (\SI{175}{\milli\electronvolt}).
The FHI98pp pseudopotential is an outlier in the set which is in line with the \(\Delta \) test results~\cite{lejaeghere-reproducibility-2016}.
We conclude that the ZPR value for ZnO agrees well with the one obtained using the new EPH module in Abinit~\cite{gonze-abinit-2020,romero-abinit-2020}.
This approach bypasses DFPT computations of the electron-phonon potential on dense q-point meshes and is thus more attractive from a computational point of view.
The value of the ZPR for ZnO-w reported in Ref.~\cite{miglio-zpr} is an outlier in the otherwise satisfactory agreement with our current results.
The remaining differences are likely to be related to substantially different pseudopotentials.
We consider the present agreement to be a strong validation of the results obtained in Ref.~\cite{miglio-zpr} as well as a thorough validation of our present VASP implementation.

\subsection{Comparison with prior work}

In this section, we compare against two publications by Karsai~\etal~\cite{karsai-one-shot} and Engel~\etal~\cite{engel-elph-paw} that calculated the band-gap ZPR using VASP, and we highlight their shortcomings.
While the approaches used in both cases employ the adiabatic Born-Oppenheimer approximation and rely on finite displacements in supercells, they differ substantially in how the ZPR is calculated.
Let us first briefly review the characteristics of each method.

In Ref.~\cite{karsai-one-shot}, the change of the band structure was determined directly from a single frozen phonon that represents a stochastically average displacement~\cite{zacharias-one-shot-2020}.
Since the band structure is calculated exactly for the frozen-phonon structure, no summation over intermediate states is required.
Additionally, this method implicitly captures contributions to the ZPR that correspond to higher-order terms in the electron-phonon interaction (anharmonicities in the ion-ion interactions are neglected, though).
The disadvantage of the supercell approach is that momentum transfers are restricted to wave vectors that are commensurate with the supercell.
In order to sample small momentum transfers, one might need to increase the supercell size, in particular for polar materials.
Presently, a simple method to accurately account for the long-range electrostatic contributions to the electron-phonon interaction does not exist in this approach, so one must rely on brute-force convergence tests and potentially huge supercells.

In Ref.~\cite{engel-elph-paw}, the ZPR was calculated from second-order perturbation theory in the harmonic approximation.
Similar to the present work, this method relies on the AHC approach and the rigid-ion approximation to calculate the DW contribution.
In contrast to the present work, the electron-phonon matrix elements were calculated using a Wannier-interpolation scheme.
While this allows for a very fine sampling of the Brillouin zone, only a comparatively small number of intermediate states has been used.
Convergence of the ZPR with respect to the number of bands is generally fairly challenging in Wannier-interpolation schemes, since conduction bands are not easily localized.
Finally, even though long-range electrostatic contributions can, in principle, be included in such an interpolation scheme, a working implementation was not available then, and these contributions are hence missing in Ref.~\cite{engel-elph-paw}.

\cref{tab:prior} lists results for the band-gap ZPR from both prior publications and the present work for selected semiconductors and insulators.
\begin{table}[!ht]
    \centering
    \caption{\label{tab:prior}
        Band-gap ZPR in \si{\milli\electronvolt}.
        Column one shows the converged ZPR from this work (\cref{tab:zpr}).
        Column two shows the adiabatic ZPR without q-point interpolation and hence without long-range electrostatic contributions.
        Columns three and four reproduce results from prior publications by Karsai~\etal{} and Engel~\etal{}, respectively, and correspond to \gridc{5} supercell calculations; in both cases, long-range electrostatic contributions are missing.
        The table includes information about whether a method employs the harmonic or adiabatic approximations, and whether it is converged with respect to the conduction-band states and q-points.
    }
    \begin{ruledtabular}
        \begin{tabular}{l r r r r}
            &
            \multicolumn{4}{c}{ZPR} \\

            &
            this work &
            this work &
            Ref.~\cite{karsai-one-shot} &
            Ref.~\cite{engel-elph-paw} \\
                            & AHC   & AHC   & one-shot  & AHC  \\
            \colrule
            adiabatic       & no    & yes   & yes       & yes   \\
            q-point conv.   & yes   & no    & no        & no    \\
            harmonic        & yes   & yes   & partly    & yes   \\
            band conv.      & yes   & yes   & yes       & no    \\

            \colrule
            AlAs-zb         & -74   & -57   & -63   & -55   \\
            AlP-zb          & -96   & -70   & -70   & -67   \\
            AlSb-zb         & -44   & -43   & -43   & -39   \\
            BN-zb           & -402  & -290  & -294  & -290  \\
            C-cd            & -323  & -320  & -320  & -337  \\
            GaN-zb          & -163  & -87   & -94   & -95   \\
            GaP-zb          & -61   & -52   & -57   & -44   \\
            Si-cd           & -58   & -54   & -65   & -54   \\
            SiC-zb          & -175  & -121  & -120  & -130  \\
        \end{tabular}
    \end{ruledtabular}
\end{table}
The reported literature values correspond to calculations using a \gridc{5} supercell and were extracted from Table~III in Ref.~\cite{engel-elph-paw}, while the first column collects the present results from \cref{tab:zpr}.
In addition, column two of \cref{tab:prior} also provides results calculated within the adiabatic AHC approximation using the present PS method, but employing supercell sizes and computational parameters that match closely those of Ref.~\cite{karsai-one-shot}.
Specifically, for column two and four, no interpolation to a dense q-point grid was performed, so the k/q-point grids are commensurate with the \gridc{5} supercell.
Convergence with respect to the number of intermediate states was obtained by summing over numerous bands as done everywhere else in this work.
We are now in a position to compare and assess the implications of the various approximations.

First, we note that the last three columns of \cref{tab:prior} are remarkably close but deviate significantly from the first column.
The last three columns have been obtained for identical supercell sizes (one-shot method) and q-point samplings.
Besides using the adiabatic approximation, all three are obviously not converged with respect to the considered momentum transfers.
For non-polar materials (C and Si) and materials with a large dielectric screening (AlSb) the resulting errors are small.
However, for the more polar materials with a small dielectric constant and large Born effective charges, the errors are not acceptable.
The errors can well reach a factor two and are typically at least \SI{50}{\percent}.
Clearly, this confirms prior observations that an accurate treatment of the long-range electrostatic interactions is required in polar materials~\cite{nery-zpr-2018}.

The second interesting assessment is the comparison of the stochastic one-shot method (third column) and the adiabatic AHC approach (second column).
The results are remarkably close, with the absolute errors hardly exceeding \SI{10}{\milli\electronvolt} (Si-cd).
The one-shot method always consistently increases the ZPR for the materials considered here.
The good agreement means that in this case the harmonic approximation in the electron-phonon interaction in combination with the rigid-ion approximation is quite well justified.
One could potentially add the corrections obtained by the one-shot method compared to the adiabatic AHC back to the q-point converged results to correct for the rigid-ion approximation and anharmonic effects in the electron-phonon interaction.

The final comparison is between column two (AHC converged with respect to intermediate conduction bands) and the AHC approach using Wannier orbitals.
The latter is not converged with respect to the intermediate states.
Clearly, the relative errors are between \SI{5}{\percent} and \SI{10}{\percent} but can approach \SI{15}{\percent} as for GaP-zb.
A systematic trend is not observed.
Generally, as shown in \cref{fig:mats-ae-vs-ps}, the convergence of the ZPR with respect to the intermediate states can be fairly erratic, with outliers often exceeding \SI{10}{\percent} of the converged ZPR.
Note that the magnitude of the ``oscillations'' also depends strongly on the used PAW potentials, as exemplified in \cref{fig:diamond-ae-vs-ps}.
Hard accurate PAW potentials (C\_h\_GW) or the AE PAW approach result in larger jumps than softer PAW potentials.
This makes the Wannier approach somewhat unpredictable.
Used with care and in combination with soft PAW potentials, it can give a fairly reliable estimate of the ZPR using little computational resources.
The downside of this approach is that one needs to be particularly careful when constructing the Wannier orbitals, and validation against converged AHC calculations will likely be required to ascertain the reliability of the results.

\section{Conclusion}
\label{sec:conclusion}

The present work determines the ZPR of the band gaps of 28 semiconductors and insulators using the PAW method.
As shown, one can derive two different expressions for the electron-phonon matrix element in the PAW method.
We have explained in detail how this is possible.
Essentially, it depends on whether the atomic derivatives are performed before or after the PAW transformation.
The PAW transformation and completeness relation can either be used at the beginning of the derivation, transforming the AE Hamiltonian to the PAW form, or alternatively after taking the derivatives of the AE Hamiltonian with respect to the ionic positions.
We have termed the former description \emph{pseudized} (PS) PAW electron-phonon matrix element and the latter \emph{all-electron} (AE) PAW electron-phonon matrix element.
The latter version describes how the electron-phonon matrix element changes as the nuclear cusp, \(Z/r \), moves against the AE orbitals.
The PS version replaces the \(Z/r \) cusp and the AE orbitals by the corresponding pseudized quantities.

On first sight, one would expect the AE description to be preferable for the determination of the ZPR, since it is fundamentally more accurate.
However, in practice, we find  a more rapid convergence of the ZPR with respect to the number of intermediate conduction-band states in the PS formulation.
Moreover, in the case of diamond, the local partial-wave expansion does not seem sufficiently complete to perform the reconstruction of the full potential in the AE formulation.
Only a reduction of the radial cutoff for the local basis yields results consistent with PS calculations.

The second issue we have addressed in the present work is the comparison between different approaches implemented in VASP.
The new reference method is based on the AHC approach, can use very dense k-point grids and can account for the energy transfer during phonon emission and absorption.
The other two considered approaches are the supercell-based one-shot method and Wannier interpolation.
The one-shot method imposes a specific phonon pattern in a supercell and determines the induced band-gap changes compared to the ground-state structure.
This approach is in some aspects more accurate than the AHC approach adopted in the present work, since it includes higher-order electron-phonon interactions and avoids the rigid-ion approximation.
Remarkably, for the materials considered here, the corrections are always below \SI{10}{\milli\electronvolt}, which we consider to be acceptable.
The disadvantage of the one-shot approach is that it neglects the energy transfer upon phonon emission and absorption, and at the often considered cell sizes it neglects important long-range contributions.
This can lead to sizable errors for polar materials.
The second method is also based on the AHC approach but replaces the reevaluation of the KS orbitals at a very dense k-point grid by a Wannier interpolation.
While this approach appears to introduce errors of about \SI{10}{\percent}, the errors can be much larger in some cases and for hard potentials.
The main source of these errors is related to the small number of conduction-band states that one can include in the Wannier approach.

Finally, we have compared our ZPR results with the results of Miglio~\etal{}~\cite{miglio-zpr} for the same set of materials.
The agreement between the two different first-principles codes is overall good, but we find a few cases where the discrepancy is approaching \SI{10}{\percent} or even \SI{20}{\percent} for materials with a very small ZPR.
We have picked one of the outliers (ZnO) and carefully reevaluated the ZPR using Abinit for several pseudopotentials.
The reevaluation greatly improved the agreement with our data.
As to why the new values using Abinit are improved compared to VASP, we have speculated that this is either related to more careful convergence tests in the present work, an unfortunate choice of pseudopotentials in the original work or recent improvements in the Abinit code.
In summary, we are confident that our present PAW results can serve as a very stringent test for other implementations.

\begin{acknowledgments}
	We express our gratitude to the authors of Ref.~\cite{miglio-zpr} for providing detailed information about their computational setup.
    Furthermore, we are grateful to Samuel Poncé for providing valuable comments on the initial preprint version of this manuscript.
\end{acknowledgments}

\appendix

\begin{widetext}

\section{Electron-phonon interaction in PAW : implementation details}
\label{app:ep-details}

\subsection{AE and PS formalisms}

If the electronic system is described by the one-electron problem,  
\begin{equation} \label{eq:app1}
    \hat{H} \ket{\psi_{\nk}}
    =
    \paren*{\frac{\hat{\mathbf{p}}^2}{2m}+\hat{v}} \ket{\psi_{\nk}}
    =
    \ev_{\nk} \ket{\psi_{\nk}}
    ,
\end{equation}
with Bloch states \(\psi_{\nk}  \) and band structure \(\ev_{\nk} \), then the electron-phonon coupling is defined as
\begin{align}
    g_{nn'\vec{k}', \nu \vec{q}}
    & =
    -\ci \frac{\ev_{\nk}-\ev_{\nkp}}{\hbar} \sum_{\kappa}
    \braket{
        \psi_{\nk} |
        -\ci \hbar \frac{\partial}{\partial \mathbf{R}_{\kappa}} |
        \psi_{\nkp}
    }
    \cdot
    \sqrt{\frac{\hbar }{2\mass_{\kappa} \omega_{\nu\mathbf{q}}}}
    \mathbf{e}_{\kappa, \nu \vec{q}}
    \Delta(\mathbf{q}+\mathbf{k}'-\mathbf{k})
    \\ & \equiv
    \sum_{\kappa} \mathbf{g}_{\nk,\nkp}^{\kappa}
    \cdot
    \sqrt{\frac{\hbar }{2\mass_{\kappa} \omega_{\nu\mathbf{q}}}}
    \mathbf{e}_{\kappa, \nu \vec{q}}
    \Delta(\mathbf{q}+\mathbf{k}'-\mathbf{k})
    .
\end{align}
In this definition, the states are normalized to one over the unit cell.
\(\kappa \) labels the atoms in the primitive cell, and \(\mathbf{e}_{\kappa, \nu\vec{q}} \) are the eigenvectors of the dynamical matrix,
\begin{equation} \label{eq:dym}
    D_{\kappa\alpha, \kappa'\alpha'}( \mathbf{q})
    =
    - \frac{1}{\sqrt{M_{\kappa} M_{\kappa'}}} \sum_{l'}
    \eul^{\ci\mathbf{q}\cdot \mathbf{R}_{l'}}
    \frac{\partial \mathbf{F}_{l' \kappa' \alpha'}}{\partial \mathbf{R}_{ \kappa \alpha}}
    ,
\end{equation}
with \(\mathbf{F}_{l' \kappa' \alpha'} \) the force on atom \(\kappa' \) in primitive cell \(l' \) in the Cartesian direction \(\alpha' \).

The matrix elements
\begin{equation} \label{eq:app2}
    \mathbf{g}_{\nk,\nkp}^{\kappa}
    =
    \braket{\psi_{\nk} | \frac{\partial \hat{H}}{\partial \mathbf{R}_{\kappa}} | \psi_{\nkp}}
    =
    \braket{\psi_{\nk} | \frac{\partial \hat{v}}{\partial \mathbf{R}_{\kappa}} | \psi_{\nkp}}
\end{equation}
have been expressed in Ref.~\cite{chaput-elphon} in terms of the orbitals and potentials used in the PAW formalism to solve \cref{eq:app1}.
In this formalism, the AE quantities are pseudized according to
\begin{align}
    \ket{\psi_{\nk}}
    & = \pawT
    \ket{\ppsi_{\nk}}
    , \\
    \paw{S} & =
    \pawT^\dagger\pawT =
    1 + \sum_{ij}\ket{\paw{p}_i}Q_{ij}\bra{\paw{p}_j}
    , \\
    \paw{H} & =
    \pawT^\dagger \hat{H} \pawT = 
    -\frac{\hbar}{2m}\nabla^2 + \paw{v}(\mathbf{r}) +
    \sum_{ij}\ket{\paw{p}_i}D_{ij}\bra{ \paw{p}_j}
    ,
\end{align}
where the \(i \) and \(j \) are compound indices to distinguish between the \(L=(n,l,m) \) momentum channels on an atom \(\kappa \), \(i=(\kappa_i,L_i) \).
Moreover, \(D_{ij} \) and \(Q_{ij} \) are localized on the atoms, and therefore are non-zero only if \(\kappa_i=\kappa_j \).
For example, \(D_{ij}=\delta_{\kappa_i,\kappa_j}D^{\kappa_i}_{L_i,L_j} \).

Expressed in terms of PS orbitals and potentials, the matrix element in \cref{eq:app2} can be written as
\begin{equation} \label{eq:g}
    \mathbf{g}_{\nk,\nkp}^{\kappa}
    =
    \mathbf{g}^{(V)}_{\nk,\nkp}(\mathbf{R}_{\kappa}) +
    \mathbf{g}^{(D)}_{\nk,\nkp}(\mathbf{R}_{\kappa}) +
    \mathbf{g}^{(P)}_{\nk,\nkp}(\mathbf{R}_{\kappa}) +
    \mathbf{g}^{(R)}_{\nk,\nkp}(\mathbf{R}_{\kappa})
    ,
\end{equation}
with
\begin{align}
    \label{eq:gv}
    \mathbf{g}^{(V)}_{\nk,\nkp} (\mathbf{R}_{ \kappa})
    & =
    \braket{
        \ppsi_{\nk} |
        \frac{d \paw{v}}{d \mathbf{R}_{\kappa}} |
        \ppsi_{\nkp}
    }
    , \\ \label{eq:gd}
    \mathbf{g}^{(D)}_{\nk,\nkp}(\mathbf{R}_{\kappa})
    & =
    \sum_{ij}
    \braket{\ppsi_{\nk} | \paw{p}_{i}}
    \frac{d  D_{ij}}{d \mathbf{R}_{\kappa}}
    \braket{\paw{p}_{j} | \ppsi_{\nkp}}
    , \\ \label{eq:gp}
    \mathbf{g}^{(P)}_{\nk,\nkp}(\mathbf{R}_{\kappa})
    & =
    \sum_{ij}
    \braket{\ppsi_{\nk} | \frac{d \paw{p}_{i}}{d \mathbf{R}_{\kappa}}}
    \Big(D_{ij} - \ev_{\nkp} Q_{ij}\Big)
    \braket{\paw{p}_{j} | \ppsi_{\nkp}}
    +
    \sum_{ij}
    \braket{\ppsi_{\nk} | \paw{p}_{i}}
    \Big(D_{ij} - \ev_{\nk} Q_{ij}\Big)
    \braket{\frac{d \paw{p}_{j}}{d \mathbf{R}_{ \kappa}} | \ppsi_{\nkp}}
    , \\ \label{eq:gr}
    \mathbf{g}^{(R)}_{\nk,\nkp}(\mathbf{R}_{\kappa})
    & =
    -(\ev_{\nk} - \ev_{\nkp}) \sum_{ij}
    \braket{\ppsi_{\nk} | \paw{p}_{i}}
    \Big(
        \braket{\phi_i|\frac{\partial \phi_j}{\partial \mathbf{R}_{\kappa}}}
        -
        \braket{\paw{\phi}_i|\frac{\partial \paw{\phi}_j}{\partial \mathbf{R}_{\kappa}}}
    \Big)
    \braket{\paw{p}_{j} | \ppsi_{\nkp}}
    .
\end{align}

In Ref.~\cite{engel-elph-paw}, an alternative approach was used to base directly the calculations on the derivatives of the overlap operator, \(\paw{S} \).
To this end, a PS electron-phonon coupling is defined as the matrix element 
\begin{align} \label{eq:app3}
    \paw{\mathbf{g}}^{\kappa}_{\nk,\nkp}
    & =
    \braket{
        \ppsi_{\nk} |
        \frac{\partial \paw{H}}{\partial \mathbf{R}_{\kappa}} - \ev_{\nkp}
        \frac{\partial \paw{S}}{\partial \mathbf{R}_{\kappa}} |
        \ppsi_{\nkp}
    } \\ & =
    \mathbf{g}^{(V)}_{\nk,\nkp}(\mathbf{R}_{\kappa}) +
    \mathbf{g}^{(D)}_{\nk,\nkp}(\mathbf{R}_{\kappa}) +
    \mathbf{g}^{(Q)}_{\nk,\nkp}(\mathbf{R}_{\kappa})
    ,
\end{align}
with 
\begin{equation}
    \mathbf{g}^{(Q)}_{\nk,\nkp}(\mathbf{R}_{\kappa})
    =
    \sum_{ij}
    \braket{\ppsi_{\nk} | \frac{d \paw{p}_{i}}{d \mathbf{R}_{\kappa}}}
    \Big(D_{ij} - \ev_{\nkp} Q_{ij}\Big)
    \braket{\paw{p}_{j} | \ppsi_{\nkp}}
    +
    \sum_{ij} \braket{\ppsi_{\nk} | \paw{p}_{i}}
    \Big(D_{ij} - \ev_{\nkp} Q_{ij}\Big)
    \braket{\frac{d \paw{p}_{j}}{d \mathbf{R}_{ \kappa}} | \ppsi_{\nkp}}
    .
\end{equation}
The use of \(\paw{\mathbf{g}}^{\kappa}_{\nk,\nkp}  \) allows to avoid the computation of the reaction matrix, \(\mathbf{g}^{ (R)}_{\nk,\nkp}( \mathbf{R}_{\kappa}) \).

\subsection{Parlinski supercell interpolation theorems for electron-phonon interactions}
\label{app:interpolation}

The electron-phonon equivalent of the Parlinski interpolation theorem for phonons~\cite{parlinski-phonon} is provided in this appendix.
In our approach, we are computing the derivatives of the potentials appearing in \cref{eq:gv,eq:gd} from finite displacements of atoms in supercell calculations (See Ref.~\cite{chaput-elphon}).
This supercell approach prevents us from accessing the quantities \(d \paw{v} / d \mathbf{R}_{\kappa} \) and \(d  D_{ij}/ d \mathbf{R}_{\kappa} \).
Indeed, having periodic boundary conditions in the supercell implies that when a finite displacement is applied to atom \(\kappa \), \(\mathbf{R}_{\kappa} \to \mathbf{R}_{\kappa} +\mathbf{u}  \), every supercell-periodic repetition of this atom is also displaced, \(\mathbf{R}_{\mathbf{L}\kappa} \to \mathbf{R}_{\mathbf{L}\kappa} +\mathbf{u}  \), with \(\mathbf{R}_{\mathbf{L}\kappa} = \mathbf{R}_{\kappa}+\mathbf{L} \), and \(\mathbf{L} \) the supercell lattice vectors.
Consequently, the quantities we obtain from finite differences are 
\begin{equation}
    \sum_{\mathbf{L}} \frac{d \paw{v}(\mathbf{r})}{d \mathbf{R}_{\mathbf{L} \kappa}}
    \text{ and }
    \sum_{\mathbf{L}} \frac{d D_{ij}}{d \mathbf{R}_{\mathbf{L} \kappa}}
    ,
\end{equation}
rather than just \(d \paw{v} / d \mathbf{R}_{\kappa} \) and \(d  D_{ij}/ d \mathbf{R}_{\kappa} \).

As shown in \cref{eq:dym}, the same aliasing problem appears in the computation of the phonon spectrum when the force constants are obtained from finite differences of forces.
In this case, instead of \(\partial \mathbf{F}_{l' \kappa' \alpha'}/\partial \mathbf{R}_{ \kappa \alpha}  \), one has \( \sum_{\mathbf{L}} \partial \mathbf{F}_{l' \kappa' \alpha'}/\partial \mathbf{R}_{ \mathbf{L} \kappa \alpha}  \), which is obtained from the supercell calculations.

The problem at hand is therefore to obtain as accurately as possible the aforementioned derivatives, \(\partial / \partial \mathbf{R}_{\kappa} \), knowing only the periodic derivatives \(\sum_{\mathbf{L}} \partial / \partial \mathbf{R}_{\mathbf{L}\kappa} \).
For phonons, by defining a proper interpolation of the dynamical matrix, Parlinski~\cite{parlinski-phonon} has shown that it is possible to obtain exact results for wave vectors \(\mathbf{q} \) that are commensurate with the supercell, \(\eul^{\ci \mathbf{q} \cdot \mathbf{L}}=1 \).
In the following for the electron-phonon coupling, we propose interpolations which also become exact for those commensurate wave vectors.

At first, for a given atom \(\kappa \) at position \(\mathbf{R}_{\kappa} \), and a given location \(\mathbf{r} \) in the supercell, we define the set of supercell lattice vectors, \(\mathbf{L} \), that minimize the distance between \(\mathbf{R}_{\kappa} \) and the image of \(\mathbf{r} \) in the neighboring supercell,
\begin{equation}
    \{\mathbf{L}\}_{\kappa, \mathbf{r}}
    =
    \argmin_{\mathbf{L}} || \mathbf{r}+\mathbf{L}-\mathbf{R}_{ \kappa} ||
    .
\end{equation}
\(\mathbf{r}+\mathbf{L} \) are the so-called minimal images of \(\mathbf{r} \) around \(\mathbf{R}_{ \kappa} \).
Then, if \(V \) is the crystal volume and \(V_S  \) the volume of the supercell, the interpolations are defined saying that the volume integration/summation is replaced by an integration/summation over the supercell volume of the same quantities, but computed with the periodic derivatives, and averaged over their minimal images around atom \(\kappa \).
This gives
\begin{align}
    \label{interpV-first}
    \mathbf{g}^{ (V)}_{\nk,\nkp} (\mathbf{R}_{ \kappa}) 
    & =
    \int_V d^3 \mathbf{r}\,
    \ppsi_{\nk}^{*}(\mathbf{r})
    \frac{d \paw{v}(\mathbf{r})}{d \mathbf{R}_{\kappa}}
    \ppsi_{\nkp}(\mathbf{r})
    \\ \label{interpV}
    & \to
    \int_{V_S} d^3 \mathbf{r}\,
    \frac{1}{|\{ \mathbf{L}\}_{ \kappa,\mathbf{r} }|}
    \sum_{\{ \mathbf{L}\}_{ \kappa, \mathbf{r} }}
    \ppsi_{\nk}^{*}(\mathbf{r}+\mathbf{L})
    \paren*{
        \sum_{\mathbf{L}'}
        \frac{d \paw{v}(\mathbf{r}+\mathbf{L})}{d \mathbf{R}_{\mathbf{L}' \kappa}}
    }
    \ppsi_{\nkp}(\mathbf{r}+\mathbf{L})
    , \\ \label{interpD-first}
    \mathbf{g}^{ (D)}_{\nk,\nkp}(\mathbf{R}_{\kappa})
    & =
    \sum_{\kappa' \in V} \sum_{LL'}
    \braket{\ppsi_{\nk} | \paw{p}_{\kappa' L}}
    \frac{d D^{\kappa'}_{ L, L'}}{d \mathbf{R}_{\kappa}}
    \braket{\paw{p}_{\kappa' L'} | \ppsi_{\nkp}}
    \\ \label{interpD}
    & \to
    \sum_{\kappa' \in V_S} \sum_{LL'}
    \frac{1}{|\{\mathbf{L}\}_{\kappa, \mathbf{R}_{\kappa'}}|}
    \sum_{\{\mathbf{L}\}_{\kappa, \mathbf{R}_{\kappa'}}}
    \braket{\ppsi_{\nk} | \paw{p}_{\mathbf{L}\kappa'L}}
    \paren*{
        \sum_{\mathbf{L}'}
        \frac{d  D^{\mathbf{L}\kappa'}_{LL'}}{d \mathbf{R}_{\mathbf{L}' \kappa}}
    }
    \braket{\paw{p}_{\mathbf{L}\kappa' L'} | \ppsi_{\nkp}}
    .
\end{align}
In \cref{interpV-first,interpD-first}, the definitions of \(\mathbf{g}^{ (V)} \) and \(\mathbf{g}^{ (D)} \) are recalled, and the interpolation is defined in \cref{interpV,interpD} with the arrow \(\to \). 

That the above interpolations becomes exact for wave vectors \(\mathbf{k} \) and \(\mathbf{k}' \) commensurate with the supercell is easily shown using the Bloch theorem and the periodicity of the potential.
Indeed, if \(V_S(\mathbf{L}') \) is the volume of the supercell with origin at \(\mathbf{L}' \), the interpolation for \(\mathbf{g}^{ (V)} \) can be written as
\begin{equation}
    \sum_{\mathbf{L}'} \int_{V_S(\mathbf{L}')} d^3\mathbf{r}\,
    \ppsi_{\nk}^{*}(\mathbf{r})
    \frac{d \paw{v}(\mathbf{r})}{d \mathbf{R}_{ \kappa}}
    \ppsi_{\nkp}(\mathbf{r})
    \frac{1}{|\{ \mathbf{L}\}_{ \kappa,\mathbf{r} }|}
    \sum_{\{\mathbf{L}\}_{\kappa, \mathbf{r}}}
    \eul^{\ci (\mathbf{k}'-\mathbf{k}) \cdot (\mathbf{L}-\mathbf{L}')}
    ,
\end{equation}
where the exponential becomes equal to 1 for commensurate wave vectors.
The interpolation of \(\mathbf{g}^{ (D)} \) can be written the same way:
\begin{equation}
    \sum_{\mathbf{L}'} \sum_{\kappa' \in V_S(\mathbf{L}')} \sum_{LL'}
    \braket{\ppsi_{\nk} | \paw{p}_{\kappa' L}}
    \frac{d D^{\kappa'}_{ L, L'}}{d \mathbf{R}_{\kappa}}
    \braket{\paw{p}_{\kappa' L'} | \ppsi_{\nkp}}
    \frac{1}{|\{ \mathbf{L}\}_{ \kappa,  \mathbf{R}_{\kappa'}}|}
    \sum_{\{\mathbf{L}\}_{\kappa, \mathbf{R}_{\kappa'}}}
    \eul^{\ci (\mathbf{k}'-\mathbf{k}) \cdot (\mathbf{L}-\mathbf{L}')}
    ,
\end{equation}
which becomes equal to \(\mathbf{g}^{(D)} \) for commensurate wave vectors.

\subsection{Treatment of the long-range part of the potential derivative}
\label{app:long-range}

In ionic systems, the potential derivative in \cref{eq:app2} may have contributions at long wavelengths, and therefore the interpolation defined in the previous section may become inaccurate.
In such a case, we may write
\begin{equation} \label{gSL}
    \mathbf{g}^{\kappa}_{\nk,\nkp}
    =
    \braket{
        \psi_{\nk} |
        \frac{\partial \hat{v}_{\kappa}^{\mathcal{S}}}{\partial \mathbf{R}_{\kappa}} |
        \psi_{\nkp}
    }
    +
    \braket{
        \psi_{\nk} |
        \frac{\partial \hat{v}_{\kappa}^{\mathcal{L}}}{\partial \mathbf{R}_{\kappa}} |
        \psi_{\nkp}
    }
    ,
\end{equation}
with 
\begin{equation}
    v_{\kappa}^{\mathcal{S}}(\mathbf{r}) =
    v(\mathbf{r}) - v_{\kappa}^{\mathcal{L}}(\mathbf{r})
    ,
\end{equation}
where \(v_{\kappa}^{\mathcal{L}} \) is a potential with a long range behavior approximately equal to the one of \(v \).
Consequently, \(v_{\kappa}^{\mathcal{S}} \) is a short-range potential to which our interpolation procedure may be applied.
In practice, we choose \(v_{\kappa}^{\mathcal{L}} \) to be the Coulomb potential due to a point charge \(Q_{\kappa} \) located at \(\mathbf{R}_{\kappa} \).
This point charge is put in a uniform background of opposite charge to keep the system neutral.
In a final step, the point charge \(Q_{\kappa} \) is promoted to be the Born effective-charge tensor, \((-e)\mathbf{Z}_{\kappa}^{\star} \), for the description of the long range part to be as accurate as possible.
Explicitly, the solution of the Poisson equation gives for \(v_{\kappa}^{\mathcal{L}} \)
\begin{equation}
    v_{\kappa}^{\mathcal{L}}(\mathbf{r})
    =
    (-e) \frac{4\pi Q_{\kappa}}{V}
    \sum_{\mathbf{q}} \sum_{\mathbf{G}\ne -\mathbf{q}}
    \frac{
        \eul^{\ci(\mathbf{q}+\mathbf{G})\cdot (\mathbf{r}-\mathbf{R}_{\kappa})}
    }{
        (\mathbf{q}+\mathbf{G}) \cdot
        \boldsymbol{\epsilon}_{\infty} \cdot
        (\mathbf{q}+\mathbf{G})
    }
    ,
\end{equation}
where \(\mathbf{G} \) are the reciprocal-lattice vectors of the crystal and \(\boldsymbol{\epsilon}_{\infty} \) is the macroscopic ion-clamped static dielectric tensor.
This gives for the second term of \cref{gSL}
\begin{equation}
    \braket{
        \psi_{\nk} |
        \frac{\partial \hat{v}_{\kappa}^{\mathcal{L}}}{\partial \mathbf{R}_{\kappa}} |
        \psi_{\nkp}
    }
    =
    (-e) \frac{4\pi Q_{\kappa}}{\Omega}
    \sum_{\mathbf{G}\ne -\mathbf{q}} \int_{\Omega} d^3 r\,
    \psi_{\nk}^{*}(\mathbf{r})
    \paren*{
        \frac{
            -\ci (\mathbf{q}+\mathbf{G})
            \eul^{\ci(\mathbf{q}+\mathbf{G}) \cdot (\mathbf{r}-\mathbf{R}_{\kappa})}
        }{
            (\mathbf{q}+\mathbf{G}) \cdot
            \boldsymbol{\epsilon}_{\infty} \cdot
            (\mathbf{q}+\mathbf{G})
        }
    }
    \psi_{\nkp}(\mathbf{r})
    ,
\end{equation}
with \(\Omega \) the volume of the primitive cell, and \(\mathbf{q}=\mathbf{k}-\mathbf{k}' \text{mod } \mathbf{G} \).
The AE orbitals, \(\psi_{\nk} \), are unknown and therefore the above quantity has to be expressed in terms of the PS orbitals, \(\ppsi_{\nk} \).
We have
\begin{equation}
    \braket{
        \psi_{\nk} |
        \eul^{\ci(\mathbf{q}+\mathbf{G})\cdot (\hat{\mathbf{r}}-\mathbf{R}_{\kappa})} |
        \psi_{\nkp}
    }
    =
    \braket{
        \ppsi_{\nk} |
        \eul^{\ci(\mathbf{q}+\mathbf{G})\cdot (\hat{\mathbf{r}}-\mathbf{R}_{\kappa})} |
        \ppsi_{\nkp}
    }
    +
    \sum_{ij}
    \braket{\ppsi_{\nk} | \paw{p}_i}
    Q_{ij}^{\kappa}(\mathbf{q}+\mathbf{G})
    \braket{\paw{p}_j | \ppsi_{\nkp}}
    ,
\end{equation}
with 
\begin{align}
    Q_{ij}^{\kappa}(\mathbf{q}+\mathbf{G})
    & =
    \braket{
        \phi_i |
        \eul^{\ci(\mathbf{q}+\mathbf{G})\cdot (\hat{\mathbf{r}}-\mathbf{R}_{\kappa})} |
        \phi_j
    }
    -
    \braket{
        \paw{\phi}_i |
        \eul^{\ci(\mathbf{q}+\mathbf{G})\cdot (\hat{\mathbf{r}}-\mathbf{R}_{\kappa})} |
        \paw{\phi}_j
    }
    \\ & =
    \int_{S_{\kappa_i}} d^3 r\,
    \paren*{
        \phi_i^{*}(\mathbf{r}) \phi_j(\mathbf{r}) -
        \paw{\phi}_i^{*}(\mathbf{r}) \paw{\phi}_j(\mathbf{r})
     }
     \eul^{\ci (\mathbf{q}+\mathbf{G}) \cdot (\mathbf{r}-\mathbf{R}_{\kappa_i})}
     \eul^{\ci (\mathbf{q}+\mathbf{G}) \cdot (\mathbf{R}_{\kappa_i}-\mathbf{R}_{\kappa})}
    .
\end{align}

\(S_{\kappa_i} \) is the atomic sphere around atom \(\kappa_i \).
However, the functions in the integral depend only on \(\mathbf{r}-\mathbf{R}_{\kappa_i} \).
Therefore, it can be evaluated in the sphere located at the origin.
Using \(\phi_i (\mathbf{r})=R_{n_i l_i}(r)S_{l_i m_i}(\hat{r})  \) (See the appendices in Ref.~\cite{chaput-elphon}) and the Rayleigh expansion formula, we obtain
\begin{equation}
    Q_{ij}^{\kappa}(\mathbf{q}+\mathbf{G})
    =
    4 \pi \eul^{\ci(\mathbf{q}+\mathbf{G}) \cdot (\mathbf{R}_{\kappa_i}-\mathbf{R}_{\kappa})}
    \sum_{l=0}^{\infty} \sum_{m=-l}^{l}
    \ci^l S_{lm}(\widehat{\mathbf{q}+\mathbf{G}}) q^l_{ij}(|\mathbf{q}+\mathbf{G}|) C_{l_i l_j m_i m_j}^{lm}
    ,
\end{equation}
where
\begin{equation}
    q^l_{ij}(|\mathbf{q}+\mathbf{G}|)
    =
    \int dr r^2\,
    \paren*{
        R_{n_i l_i}(r) R_{n_j l_j}(r) -
        \paw{R}_{n_i l_i}(r) \paw{R}_{n_j l_j}(r)
    }
    j_l(|\mathbf{q}+\mathbf{G}|r)
    ,
\end{equation}
with the spherical Bessel functions \(j_l \) and the Gaunt coefficients, 
\begin{equation}
    C_{l_i l_j m_i m_j}^{lm}
    =
    \int d\hat{r}\, S_{l_i m_i}(\hat{r}) S_{l m}(\hat{r}) S_{l_j m_j}(\hat{r})
    .
\end{equation}
This gives
\begin{equation}
    \begin{split}
        \braket{
            \psi_{\nk} |
            \frac{\partial \hat{v}_{\kappa}^{\mathcal{L}}}{\partial \mathbf{R}_{\kappa}} |
            \psi_{\nkp}
        }
        = (-e) &
        \frac{4\pi Q_{\kappa}}{\Omega}
        \sum_{\mathbf{G}\ne -\mathbf{q}} \int_{\Omega} d^3 r\,
        \ppsi_{\nk}^{*}(\mathbf{r})
        \paren*{
            \frac{
                -\ci (\mathbf{q}+\mathbf{G})
                \eul^{\ci (\mathbf{q}+\mathbf{G}) \cdot (\mathbf{r}-\mathbf{R}_{\kappa})}
            }{
                (\mathbf{q}+\mathbf{G}) \cdot
                \boldsymbol{\epsilon}_{\infty} \cdot
                (\mathbf{q}+\mathbf{G})
            }
        }
        \ppsi_{\nkp}(\mathbf{r})
        \\ {}+ (-e) &
        \frac{4\pi Q_{\kappa}}{\Omega}
        \sum_{\mathbf{G}\ne -\mathbf{q}} \sum_{ij}
        \frac{
            -\ci (\mathbf{q}+\mathbf{G})
        }{
            (\mathbf{q}+\mathbf{G}) \cdot
            \boldsymbol{\epsilon}_{\infty} \cdot
            (\mathbf{q}+\mathbf{G})
        }
        \braket{\ppsi_{\nk} | \paw{p}_i}
        Q_{ij}^{\kappa}(\mathbf{q}+\mathbf{G})
        \braket{\paw{p}_j | \ppsi_{\nkp}}
        .
    \end{split}
\end{equation}

To compute \(\braket{\psi_{\nk} | \frac{\partial \hat{v}_{\kappa}^{\mathcal{S}}}{\partial \mathbf{R}_{\kappa}} | \psi_{\nkp}} \), the interpolation procedure defined in the previous section is used.
The quantities \(\sum_{\mathbf{L}'} d \paw{v}/ d \mathbf{R}_{\mathbf{L}'\kappa} \) and \(\sum_{\mathbf{L}'} d D_{ij}/d \mathbf{R}_{\mathbf{L}'\kappa} \), which appear in \cref{interpV,interpD}, are the change of the PAW potential when an atom \(\kappa \) is moved in every periodic repetition of the supercell with lattice vectors \(\mathbf{L} \).
When the Poisson equation is solved with such periodic boundary conditions, its solution can be written as
\begin{equation}
    \sum_{\mathbf{L}}
    \frac{d v^{\mathcal{L}}(\mathbf{r})}{d \mathbf{R}_{\mathbf{L}\kappa}}
    =
    (-e) \frac{d}{d \mathbf{R}_{\kappa}} \frac{4\pi Q_{\kappa}}{V_S}
    \sum_{\mathbf{K} \ne 0}
    \frac{
        \eul^{\ci \mathbf{K} \cdot (\mathbf{r}-\mathbf{R}_{\kappa})}
    }{
        \mathbf{K} \cdot
        \boldsymbol{\epsilon}_{\infty} \cdot
        \mathbf{K}
    }
    ,
\end{equation}
where \(\mathbf{K} \) are the reciprocal-lattice vectors of the supercell.
To obtain the PAW representation of this potential, once again we write
\begin{equation}
    \braket{
        \psi_{\nk} |
        \eul^{\ci\mathbf{K}\cdot (\mathbf{r}-\mathbf{R}_{\kappa})} |
        \psi_{\nkp}
    }
    =
    \braket{
        \ppsi_{\nk} |
        \eul^{\ci\mathbf{K}\cdot (\mathbf{r}-\mathbf{R}_{\kappa})} |
        \ppsi_{\nkp}
    }
    +
    \sum_{ij}
    \braket{\ppsi_{\nk} | \paw{p}_i}
    Q_{ij}^{\kappa}(\mathbf{K})
    \braket{\paw{p}_j | \ppsi_{\nkp}}
    .
\end{equation}
Therefore,
\begin{equation}
    \begin{split}
        \braket{
            \psi_{\nk} |
            \sum_{\mathbf{L}} \frac{d \hat{v}^{\mathcal{L}}}{d \mathbf{R}_{\mathbf{L}\kappa}} |
            \psi_{\nkp}
        }
        & =
        \braket{
            \ppsi_{\nk} |
            (-e) \frac{4\pi Q_{\kappa}}{V_S}
            \sum_{\mathbf{K} \ne 0}
            \frac{
                -\ci \mathbf{K}
            }{
                \mathbf{K} \cdot
                \boldsymbol{\epsilon}_{\infty} \cdot
                \mathbf{K}
            }
            \eul^{\ci \mathbf{K} \cdot (\hat{\mathbf{r}} - \mathbf{R}_{\kappa})} |
            \ppsi_{\nkp}
        }
        \\
        & + \sum_{ij}
        \braket{\ppsi_{\nk} | \paw{p}_i}
        (-e) \frac{4\pi Q_{\kappa}}{V_S}
        \sum_{\mathbf{K} \ne 0}
        \frac{
            -\ci \mathbf{K}
        }{
            \mathbf{K} \cdot
            \boldsymbol{\epsilon}_{\infty} \cdot
            \mathbf{K}
        }
        Q_{ij}^{\kappa}(\mathbf{K})
        \braket{\paw{p}_j | \ppsi_{\nkp}}
        .
    \end{split}
\end{equation}
Comparing with \cref{eq:gv,eq:gd}, this last equation shows that we can use the equation of the previous section to perform the interpolation procedure by making the substitutions
\begin{align}
    \sum_{\mathbf{L}}
    \frac{d \paw{v}(\mathbf{r})}{d \mathbf{R}_{\mathbf{L}\kappa}}
    & \longleftarrow
    \sum_{\mathbf{L}}
    \frac{d \paw{v}(\mathbf{r})}{d \mathbf{R}_{\mathbf{L}\kappa}}
    -
    \frac{4\pi e^2 \mathbf{Z}^{\star \top}_\kappa}{V_S}
    \sum_{\mathbf{K} \ne 0}
    \frac{
        -\ci \mathbf{K}
    }{
        \mathbf{K} \cdot
        \boldsymbol{\epsilon}_{\infty} \cdot
        \mathbf{K}
    }
    \eul^{\ci \mathbf{K}\cdot (\mathbf{r}-\mathbf{R}_{\kappa})}
    , \\
    \sum_{\mathbf{L}}
    \frac{d D_{ij}}{d \mathbf{R}_{\mathbf{L}\kappa}}
    & \longleftarrow
    \sum_{\mathbf{L}}
    \frac{d D_{ij}}{d \mathbf{R}_{\mathbf{L}\kappa}}
    -
    \frac{4\pi e^2 \mathbf{Z}^{\star \top}_\kappa}{V_S}
    \sum_{\mathbf{K} \ne 0}
    \frac{
        -\ci \mathbf{K}
    }{
        \mathbf{K} \cdot
        \boldsymbol{\epsilon}_{\infty} \cdot
        \mathbf{K}
    }
    Q_{ij}^{\kappa}(\mathbf{K})
    .
\end{align}

\end{widetext}

\section{AE and PS methods in the adiabatic approximation}
\label{app:ae-ps-equiv-proof}

In this section, we show the formal equivalence of the band-structure renormalization between the AE and PS formulations in the adiabatic limit by means of an algebraic proof.
For simplicity's sake, we focus on the ZPR but the proof is equally valid at finite temperature.
In addition, we assume that the band structure is non-degenerate everywhere.
While not strictly necessary, this allows for a clean removal of the \(\ci \delta \) term from the denominator of the self-energy expression.
Furthermore, in this proof, we do \emph{not} employ the rigid-ion approximation, which means that the DW contribution is exact in second-order perturbation theory.
Under these assumptions, the FM contribution to the adiabatic ZPR reads
\begin{align} \label{eq:app:zpr-fm}
    \text{ZPR}^\text{FM,a}_\nk
    & =
    \bzint \kint{q} \sum_{\nu}
    \gamma^{\text{FM}}_{\nk,\nu \vec{q}}
    , \\ \label{eq:app:zpr-fm-gamma-ae}
    \gamma^{\text{FM}}_{\nk,\nu \vec{q}}
    & \equiv
    \sum_{m}'
    \frac{
        \abs{g_{mn \vec{k}, \nu \vec{q}}}^2
    }{
        \ev_\nk - \ev_{m \vec{k} + \vec{q}}
    }
    ,
\end{align}
where the prime above the sum indicates that the case \((m \vec{k} + \vec{q}) = (\nk) \) is excluded.
The DW contribution takes the form
\begin{align} \label{eq:app:zpr-dw}
    \text{ZPR}^\text{DW,a}_\nk
    & =
    \bzint \kint{q} \sum_{\nu}
    \gamma^{\text{DW}}_{\nk,\nu \vec{q}}
    , \\ \label{eq:app:zpr-dw-gamma-ae}
    \gamma^{\text{DW}}_{\nk,\nu \vec{q}}
    & \equiv
    \frac{1}{2}
    \braket{\psi_\nk | \partial_{\nu \vec{q}} \partial^*_{\nu \vec{q}} \hat{H} | \psi_\nk}
    ,
\end{align}
where \(\partial^*_{\nu \vec{q}} = \partial_{\nu, -\vec{q}} \).

In the PS formulation, the terms \(\gamma^{\text{FM}}_{\nk,\nu \vec{q}} \) and \(\gamma^{\text{DW}}_{\nk,\nu \vec{q}} \) are replaced by their respective PS equivalents:
\begin{multline} \label{eq:app:zpr-fm-gamma-ps}
    \paw{\gamma}^{\text{FM}}_{\nk,\nu \vec{q}}
    \equiv
    \sum_{m}'
    \frac{
        \abs{\paw{g}_{mn \vec{k}, \nu \vec{q}}}^2
    }{
        \ev_\nk - \ev_{m \vec{k} + \vec{q}}
    }
    \\ -
    \paw{g}_{nn \vec{k}, \nu \vec{0}}
    \braket{\ppsi_\nk | \partial_{\nu \vec{0}} \paw{S} | \ppsi_\nk}
    \delta\paren{\vec{q}}
\end{multline}
and
\begin{equation} \label{eq:app:zpr-dw-gamma-ps}
    \paw{\gamma}^{\text{DW}}_{\nk,\nu \vec{q}}
    \equiv
    \frac{1}{2}
    \braket{
        \psi_\nk |
        \partial_{\nu \vec{q}} \partial^*_{\nu \vec{q}} \paw{H}
        - \ev_\nk \partial_{\nu \vec{q}} \partial^*_{\nu \vec{q}} \paw{S} |
        \psi_\nk
    }
    ,
\end{equation}
which yields the ZPR equations reported in Ref.~\cite{engel-elph-paw}.
Our goal is to show that \(\gamma^{\text{FM}}_{\nk,\nu \vec{q}} - \paw{\gamma}^{\text{FM}}_{\nk,\nu \vec{q}} = \paw{\gamma}^{\text{DW}}_{\nk,\nu \vec{q}} - \gamma^{\text{DW}}_{\nk,\nu \vec{q}} \) when integrated over the first Brillouin zone, which proves that both formulations yield identical results for the adiabatic ZPR.

Let us begin by transforming the AE electron-phonon matrix elements in \cref{eq:app:zpr-fm-gamma-ae} to the PS formulation using \cref{eq:g-ae-vs-ps}.
To keep track of the different resulting terms, they are divided into three different contributions:
\begin{align}
    \gamma^{\text{FM}}_{\nk,\nu \vec{q}}
    & =
    \gamma^{\text{FM,a}}_{\nk,\nu \vec{q}} +
    \gamma^{\text{FM,b}}_{\nk,\nu \vec{q}} +
    \gamma^{\text{FM,c}}_{\nk,\nu \vec{q}}
    , \\ \label{eq:app:fm-a-def}
    \gamma^{\text{FM,a}}_{\nk,\nu \vec{q}}
    & \equiv
    \sum_{m}'
    \frac{
        \abs{\paw{g}_{mn \vec{k}, \nu \vec{q}}}^2
    }{
        \ev_\nk - \ev_{m \vec{k} + \vec{q}}
    }
    , \\
    \gamma^{\text{FM,b}}_{\nk,\nu \vec{q}}
    & \equiv
    \sum_{m}'
    \paw{g}^*_{mn \vec{k}, \nu \vec{q}}
    \paw{t}_{mn \vec{k}, \nu \vec{q}}
    + \text{c.c.}
    , \\
    \gamma^{\text{FM,c}}_{\nk,\nu \vec{q}}
    & \equiv
    \sum_{m}'
    \paren{\ev_\nk - \ev_{m \vec{k} + \vec{q}}}
    \abs*{\paw{t}_{mn \vec{k}, \nu \vec{q}}}^2
    , \\
    \paw{t}_{mn \vec{k}, \nu \vec{q}}
    & \equiv
    \braket{
        \ppsi_{m \vec{k} + \vec{q}} |
        \pawT^\dagger \partial_{\nu \vec{q}} \pawT |
        \ppsi_\nk
    }
    .
\end{align}
The term \(\gamma^{\text{FM,a}}_{\nk,\nu \vec{q}} \) is already identical to the first line of \cref{eq:app:zpr-fm-gamma-ps}.
Next, let us add and subtract the \((m \vec{q}) = (n \vec{0}) \) contribution to and from \(\gamma^{\text{FM,b}}_{\nk,\nu \vec{q}} \) to complete the sum:
\begin{multline} \label{eq:app:fm-b-intermed-1}
    \gamma^{\text{FM,b}}_{\nk,\nu \vec{q}}
    =
    \sum_{m} \left[
        \paw{g}^*_{mn \vec{k}, \nu \vec{q}}
        \paw{t}_{mn \vec{k}, \nu \vec{q}}
        + \text{c.c.}
    \right]
    \\ -
    \paw{g}^*_{nn \vec{k}, \nu \vec{0}}
    \paw{t}_{nn \vec{k}, \nu \vec{0}}
    \delta\paren{\vec{q}}
    - \text{c.c.}
    .
\end{multline}
Since \(\partial_{\nu \vec{0}} = \partial^*_{\nu \vec{0}} \), we know that \(\paw{g}_{nn \vec{k}, \nu \vec{0}} \) is real, which lets us further modify the terms appearing on the second line of \cref{eq:app:fm-b-intermed-1}:
\begin{align*}
    &
    \paw{g}^*_{nn \vec{k}, \nu \vec{0}}
    \paw{t}_{nn \vec{k}, \nu \vec{0}}
    +
    \paw{t}^*_{nn \vec{k}, \nu \vec{0}}
    \paw{g}_{nn \vec{k}, \nu \vec{0}}
    \\ ={} &
    \paw{g}_{nn \vec{k}, \nu \vec{0}}
    \left[
        \paw{t}_{nn \vec{k}, \nu \vec{0}} +
        \paw{t}^*_{nn \vec{k}, \nu \vec{0}}
    \right]
    \\ ={} &
    \paw{g}_{nn \vec{k}, \nu \vec{0}}
    \braket{\ppsi_\nk | \partial_{\nu \vec{0}} \paw{S} | \ppsi_\nk}
    ,
\end{align*}
where we used the product rule of differentiation to retrieve the PAW overlap operator.
Therefore,
\begin{multline} \label{eq:app:fm-b-intermed-2}
    \gamma^{\text{FM,b}}_{\nk,\nu \vec{q}}
    =
    \sum_{m} \left[
        \paw{g}^*_{mn \vec{k}, \nu \vec{q}}
        \paw{t}_{mn \vec{k}, \nu \vec{q}}
        + \text{c.c.}
    \right]
    \\ -
    \paw{g}_{nn \vec{k}, \nu \vec{0}}
    \braket{\ppsi_\nk | \partial_{\nu \vec{0}} \paw{S} | \ppsi_\nk}
    \delta\paren{\vec{q}}
    .
\end{multline}
Summing \cref{eq:app:fm-a-def} and the second line of \cref{eq:app:fm-b-intermed-2} yields exactly \(\paw{\gamma}^{\text{FM}}_{\nk,\nu \vec{q}} \), so the difference of the AE and PS FM contributions is simply the entire remainder:
\begin{multline} \label{eq:app:fm-diff}
    \gamma^{\text{FM}}_{\nk,\nu \vec{q}}
    - \paw{\gamma}^{\text{FM}}_{\nk,\nu \vec{q}}
    = \\
    \sum_{m}
    \Bigl[
        \paren{\ev_\nk - \ev_{m \vec{k} + \vec{q}}}
        \abs*{\paw{t}_{mn \vec{k}, \nu \vec{q}}}^2
        \\ +
        \paw{g}^*_{mn \vec{k}, \nu \vec{q}}
        \paw{t}_{mn \vec{k}, \nu \vec{q}}
        +
        \paw{t}^*_{mn \vec{k}, \nu \vec{q}}
        \paw{g}_{mn \vec{k}, \nu \vec{q}}
    \Bigr]
    .
\end{multline}
Notice how \cref{eq:app:fm-diff} now contains a sum over all intermediate states, \(m \).
This is possible since the case \((m \vec{q}) = (n \vec{0}) \) does not contribute in \(\gamma^{\text{FM,c}}_{\nk,\nu \vec{q}} \).

To show the difference between the AE and PS DW contributions, it is easier to begin in the PS formulation by expanding \cref{eq:app:zpr-dw-gamma-ps} in terms of the PAW transformation and then applying the differential operators using the product rule.
This way, the second derivative of the PAW Hamiltonian contributes 9 terms while the second derivative of the PAW overlap contributes 4.
To keep track of all these terms, we once again group them into different contributions:
\begin{widetext}
    \begin{align}
        \label{eq:app:dw-contrib-sum}
        \paw{\gamma}^{\text{DW}}_{\nk,\nu \vec{q}}
        & =
        \gamma^{\text{DW}}_{\nk,\nu \vec{q}} +
        \paw{\gamma}^{\text{DW,a}}_{\nk,\nu \vec{q}} +
        \paw{\gamma}^{\text{DW,b}}_{\nk,\nu \vec{q}} +
        \paw{\gamma}^{\text{DW,c}}_{\nk,\nu \vec{q}}
        , \\
        \paw{\gamma}^{\text{DW,a}}_{\nk,\nu \vec{q}}
        & \equiv
        \frac{1}{2}
        \Braket{
            \ppsi_\nk |
            \partial_{\nu \vec{q}} \partial^*_{\nu \vec{q}} \pawT^\dagger \hat{H} \pawT
            -
            \ev_\nk
            \partial_{\nu \vec{q}} \partial^*_{\nu \vec{q}} \pawT^\dagger \pawT
            | \ppsi_\nk
        }
        + \text{c.c.}
        , \\ \label{eq:app:dw-b-def}
        \paw{\gamma}^{\text{DW,b}}_{\nk,\nu \vec{q}}
        & \equiv
        \frac{1}{2}
        \Braket{
            \ppsi_\nk |
            \partial^*_{\nu \vec{q}} \pawT^\dagger \hat{H} \partial_{\nu \vec{q}} \pawT
            +
            \partial_{\nu \vec{q}} \pawT^\dagger \hat{H} \partial^*_{\nu \vec{q}} \pawT
            -
            \ev_\nk
            \partial^*_{\nu \vec{q}} \pawT^\dagger \partial_{\nu \vec{q}} \pawT
            -
            \ev_\nk
            \partial_{\nu \vec{q}} \pawT^\dagger \partial^*_{\nu \vec{q}} \pawT
            | \ppsi_\nk
        }
        , \\ \label{eq:app:dw-c-def}
        \paw{\gamma}^{\text{DW,c}}_{\nk,\nu \vec{q}}
        & \equiv
        \frac{1}{2}
        \Braket{
            \ppsi_\nk |
            \pawT^\dagger \partial^*_{\nu \vec{q}} \hat{H} \partial_{\nu \vec{q}} \pawT
            +
            \pawT^\dagger \partial_{\nu \vec{q}} \hat{H} \partial^*_{\nu \vec{q}} \pawT
            +
            \partial^*_{\nu \vec{q}} \pawT^\dagger \partial_{\nu \vec{q}} \hat{H} \pawT
            +
            \partial_{\nu \vec{q}} \pawT^\dagger \partial^*_{\nu \vec{q}} \hat{H} \pawT
            | \ppsi_\nk
        }
        .
    \end{align}
\end{widetext}
The first term on \cref{eq:app:dw-contrib-sum} is already the AE DW contribution.
Using the KS equations, \(\hat{H} \ket{\psi_\nk} = \ev_\nk \ket{\psi_\nk} \), it is quite easy to show that \(\paw{\gamma}^{\text{DW,a}}_{\nk,\nu \vec{q}} \) is, in fact, zero.
\cref{eq:app:dw-b-def,eq:app:dw-c-def} consist of pairs of terms which are invariant under a sign change of \(\vec{q} \).
For example, the first term in \cref{eq:app:dw-b-def} maps onto the second term:
\begin{equation*}
    \partial^*_{\nu, -\vec{q}} \pawT^\dagger \hat{H} \partial_{\nu, -\vec{q}} \pawT
    =
    \partial_{\nu \vec{q}} \pawT^\dagger \hat{H} \partial^*_{\nu \vec{q}} \pawT
    .
\end{equation*}
Since the domain of integration is the first Brillouin zone and therefore inversion symmetric around the origin, the sign of the integration variable, \(\vec{q} \), can be swapped without affecting the result.
Hence, by redefining \(\vec{q} \to -\vec{q} \) for only one term of each pair mentioned earlier, we can simplify the contributions from \cref{eq:app:dw-b-def},
\begin{multline} \label{eq:app:dw-b-intermed}
    \bzint \kint{q}
    \paw{\gamma}^{\text{DW,b}}_{\nk,\nu \vec{q}}
    = \\
    \bzint \kint{q}
    \braket{
        \ppsi_\nk |
        \partial^*_{\nu \vec{q}} \pawT^\dagger \hat{H} \partial_{\nu \vec{q}} \pawT |
        \ppsi_\nk
    }
    \\ - \ev_\nk
    \bzint \kint{q}
    \braket{
        \ppsi_\nk |
        \partial^*_{\nu \vec{q}} \pawT^\dagger \partial_{\nu \vec{q}} \pawT |
        \ppsi_\nk
    }
    ,
\end{multline}
and from \cref{eq:app:dw-c-def},
\begin{multline} \label{eq:app:dw-c-intermed-1}
    \bzint \kint{q}
    \paw{\gamma}^{\text{DW,c}}_{\nk,\nu \vec{q}}
    = \\
    \bzint \kint{q}
    \braket{
        \ppsi_\nk |
        \pawT^\dagger \partial^*_{\nu \vec{q}} \hat{H} \partial_{\nu \vec{q}} \pawT |
        \ppsi_\nk
    }
    \\ +
    \bzint \kint{q}
    \braket{
        \ppsi_\nk |
        \partial^*_{\nu \vec{q}} \pawT^\dagger \partial_{\nu \vec{q}} \hat{H} \pawT |
        \ppsi_\nk
    }
    .
\end{multline}
Each of the matrix elements appearing in \cref{eq:app:dw-b-intermed,eq:app:dw-c-intermed-1} consists of a product of two first-order derivatives.
In order to separate them, we insert the complete basis of PS Bloch states,
\begin{equation*}
    \id =
    \bzint \kint{k} \sum_m \pawT \ket{\ppsi_{m \vec{k}}} \bra{\ppsi_{m \vec{k}}} \pawT^\dagger
    ,
\end{equation*}
in-between the operators.
Due to the Bloch theorem, it is sufficient to only involve states at \(\vec{k} + \vec{q} \) to complete the basis in this case.
Let us demonstrate this procedure for the first matrix element in \cref{eq:app:dw-b-intermed}:
\begin{widetext}
    \begin{align*}
        \partial^*_{\nu \vec{q}} \pawT^\dagger \hat{H} \partial_{\nu \vec{q}} \pawT
        & =
        \partial^*_{\nu \vec{q}} \pawT^\dagger
        \sum_m \left[
            \pawT \ket{\ppsi_{m \vec{k} + \vec{q}}}
            \bra{\ppsi_{m \vec{k} + \vec{q}}} \pawT^\dagger
        \right]
        \hat{H}
        \sum_{m'} \left[
            \pawT \ket{\ppsi_{m' \vec{k} + \vec{q}}}
            \bra{\ppsi_{m' \vec{k} + \vec{q}}} \pawT^\dagger
        \right]
        \partial_{\nu \vec{q}} \pawT
        \\ & =
        \sum_{mm'}
        \partial^*_{\nu \vec{q}} \pawT^\dagger \pawT
        \ket{\ppsi_{m \vec{k} + \vec{q}}}
        \underbrace{
            \braket{
                \ppsi_{m \vec{k} + \vec{q}} |
                \paw{H} |
                \ppsi_{m' \vec{k} + \vec{q}}
            }
        }_{
            \ev_{m \vec{k} + \vec{q}} \delta_{mm'}
        }
        \bra{\ppsi_{m' \vec{k} + \vec{q}}}
        \pawT^\dagger \partial_{\nu \vec{q}} \pawT
        \\ & =
        \sum_m \ev_{m \vec{k} + \vec{q}}
        \partial^*_{\nu \vec{q}} \pawT^\dagger \pawT
        \ket{\ppsi_{m \vec{k} + \vec{q}}}
        \bra{\ppsi_{m \vec{k} + \vec{q}}}
        \pawT^\dagger \partial_{\nu \vec{q}} \pawT
        .
    \end{align*}
    Using this trick to separate the derivatives, \cref{eq:app:dw-b-intermed} becomes
    \begin{equation}
        \begin{split}
            \bzint \kint{q}
            \paw{\gamma}^{\text{DW,b}}_{\nk,\nu \vec{q}}
            & =
            - \bzint \kint{q} \sum_m \paren{\ev_\nk - \ev_{m \vec{k} + \vec{q}}}
            \braket{
                \ppsi_\nk |
                \partial^*_{\nu \vec{q}} \pawT^\dagger \pawT |
                \ppsi_{m \vec{k} + \vec{q}}
            }
            \braket{
                \ppsi_{m \vec{k} + \vec{q}} |
                \pawT^\dagger \partial_{\nu \vec{q}} \pawT |
                \ppsi_\nk
            }
            \\ & =
            - \bzint \kint{q} \sum_m \paren{\ev_\nk - \ev_{m \vec{k} + \vec{q}}}
            \abs*{\paw{t}_{mn \vec{k}, \nu \vec{q}}}^2
            .
        \end{split}        
    \end{equation}
    Similarly, \cref{eq:app:dw-c-intermed-1} becomes
    \begin{equation} \label{eq:app:dw-c-intermed-2}
        \begin{split}
            \bzint \kint{q}
            \paw{\gamma}^{\text{DW,c}}_{\nk,\nu \vec{q}}
            & 
            \!\begin{multlined}[t]
                =
                \bzint \kint{q} \sum_m
                \Bigl[
                    \braket{
                        \ppsi_\nk |
                        \pawT^\dagger \partial^*_{\nu \vec{q}} \hat{H} \pawT |
                        \ppsi_{m \vec{k} + \vec{q}}
                    }
                    \braket{
                        \ppsi_{m \vec{k} + \vec{q}} |
                        \pawT^\dagger \partial_{\nu \vec{q}} \pawT |
                        \ppsi_\nk
                    }
                    \\ +
                    \braket{
                        \ppsi_\nk |
                        \partial^*_{\nu \vec{q}} \pawT^\dagger  \pawT |
                        \ppsi_{m \vec{k} + \vec{q}}
                    }
                    \braket{
                        \ppsi_{m \vec{k} + \vec{q}} |
                        \pawT^\dagger \partial_{\nu \vec{q}} \hat{H} \pawT |
                        \ppsi_\nk
                    }
                \Bigr]
            \end{multlined}
            \\ & =
            \bzint \kint{q} \sum_m
            \Bigl[
                g^*_{mn \vec{k}, \nu \vec{q}}
                t_{mn \vec{k}, \nu \vec{q}}
                +
                t^*_{mn \vec{k}, \nu \vec{q}}
                g_{mn \vec{k}, \nu \vec{q}}
            \Bigr]
            .
        \end{split}
    \end{equation}
\end{widetext}
The final step involves transforming the AE matrix elements in \cref{eq:app:dw-c-intermed-2} using \cref{eq:g-ae-vs-ps} to the PS representation:
\begin{equation}
    \begin{split}
        & g^*_{mn \vec{k}, \nu \vec{q}}
        t_{mn \vec{k}, \nu \vec{q}}
        +
        t^*_{mn \vec{k}, \nu \vec{q}}
        g_{mn \vec{k}, \nu \vec{q}}
        \\ ={} &
        \paw{g}^*_{mn \vec{k}, \nu \vec{q}}
        t_{mn \vec{k}, \nu \vec{q}}
        +
        t^*_{mn \vec{k}, \nu \vec{q}}
        \paw{g}_{mn \vec{k}, \nu \vec{q}}
        \\
        & + 2
        \paren{\ev_\nk - \ev_{m \vec{k} + \vec{q}}}
        \abs*{\paw{t}_{mn \vec{k}, \nu \vec{q}}}^2
        .
    \end{split}
\end{equation}
This allows us to state the difference between the integrated AE and PS DW contributions:
\begin{multline} \label{eq:app:dw-diff}
    \bzint \kint{q} \left[
        \paw{\gamma}^{\text{DW}}_{\nk,\nu \vec{q}}
        -
        \gamma^{\text{DW}}_{\nk,\nu \vec{q}}
    \right]
    = \\
    \bzint \kint{q} \sum_m \Bigl[
        \paren{\ev_\nk - \ev_{m \vec{k} + \vec{q}}}
        \abs*{\paw{t}_{mn \vec{k}, \nu \vec{q}}}^2
        \\ +
        \paw{g}^*_{mn \vec{k}, \nu \vec{q}}
        \paw{t}_{mn \vec{k}, \nu \vec{q}}
        +
        \paw{t}^*_{mn \vec{k}, \nu \vec{q}}
        \paw{g}_{mn \vec{k}, \nu \vec{q}}
    \Bigr]
    .
\end{multline}
Finally, we are able to compare \cref{eq:app:dw-diff} against \cref{eq:app:fm-diff} and find that
\begin{multline}
    \bzint \kint{q} \left[
        \gamma^{\text{FM}}_{\nk,\nu \vec{q}}
        - \paw{\gamma}^{\text{FM}}_{\nk,\nu \vec{q}}
    \right]
    \\ =
    \bzint \kint{q} \left[
        \paw{\gamma}^{\text{DW}}_{\nk,\nu \vec{q}}
        - \gamma^{\text{DW}}_{\nk,\nu \vec{q}}
    \right]
    ,
\end{multline}
which, of course, implies that the AE and PS ZPR equations are identical.

Note that this equivalence between the AE and PS formulations is no longer valid once the rigid-ion approximation is introduced.
This approximation is used to calculate the second-order DW contribution with greater computational efficiency.
However, the way this approximation is employed in the PS case detailed in Ref.~\cite{engel-elph-paw} is different from how it is employed in the AE case, in the sense that the resulting ZPR expressions are no longer formally equivalent.
Due to the subtlety of the issue and its potential impact on calculations, it requires further investigation in the future.

\bibliographystyle{apsrev4-2}
\bibliography{references}

\end{document}